\documentclass[aps,prd,a4paper,preprintnumbers,tightenlines,superscriptaddress,twocolumn,floatfix,showpacs,final]{revtex4}
\usepackage{graphicx}
\usepackage{amsmath}
\usepackage{amssymb}
\usepackage{subfigure}
\usepackage{color}
\newcommand{\vev}{\; {\!\mathrm{VEV}}}
\newcommand{\be}{\begin{eqnarray}}
\newcommand{\ee}{\end{eqnarray}}
\newcommand{\mpl}{{M_{\rm {pl}}}}
\newcommand{\pp}{\; \!\!\!\left(\phi\right)}
\newcommand{\dd}{\, {\rm d}}
\newcommand{\gsim}{\;\mbox{\raisebox{-0.5ex}{$\stackrel{>}{\scriptstyle{\sim}}$}}\;}
\newcommand{\lsim}{\;\mbox{\raisebox{-0.5ex}{$\stackrel{<}{\scriptstyle{\sim}}$}}\;}

\begin{document}
\title{Modified Gravity Makes Galaxies Brighter}
\author{Anne-Christine Davis}
\author{Eugene A. Lim}
\author{Jeremy Sakstein}
\author{Douglas J. Shaw}
\affiliation{Department of Applied Mathematics and Theoretical Physics, Centre for Mathematical Sciences, Cambridge CB3 0WA,
United Kingdom}

\begin{abstract}

We investigate the effect of modified gravity with screening mechanisms, such as the chameleon or symmetron models, upon the structure of main sequence stars. We find that unscreened stars can be significantly more luminous and ephemeral than their screened doppelgangers. By embedding these stars into dwarf galaxies, which can be unscreened for values of the parameters not yet ruled out observationally, we show that the cumulative effect of their increased luminosity can enhance the total galactic luminosity. We estimate this enhancement and find that it can be considerable given model parameters that are still under experimental scrutiny. By looking for systematic offsets between screened dwarf galaxies in clusters and unscreened galaxies in voids, these effects could form the basis of an independent observational test that can potentially lower the current experimental bounds on the model independent parameters of these theories by and order of magnitude or more.  

\end{abstract}
\maketitle

\section{Introduction}

Over the last decade or so, there has been a plethora of evidence that the universe is accelerating (see, for example \cite{Copeland:2006wr} and references therein). It is well known that all observed matter that interacts via gravity as described by general relativity (GR) feels an attractive force. This should cause the universe to decelerate, or at least remain stationary. Many of the attempts to explain this phenomena have tried either to introduce some exotic form of matter in the form of dark energy or to propose a modified theory of gravity. Many of these new forms of matter take the form of a scalar field and Weinberg \cite{Weinberg:1965rz} showed that any modification to general relativity necessarily introduces a new degree of freedom and so these two different approaches are, on some level, equivalent. One would generically expect any new scalar degree of freedom to couple to ordinary matter, which invariably leads to the introduction of an additional, or \textit{fifth}, force which alters the strength of the gravitational attraction. These forces are absent in general relativity, whose predictions have been tested to high precision over the years in both the laboratory and the solar system (see \cite{Will:2004nx} for some examples). In particular, these tests have placed very stringent constraints on any additional forces. Either they must be short ranged ($\lsim$ 0.1 mm) or their strength must be weak compared to gravity. One is then naively led to conclude that these experiments rule out any theories of modified gravity that predict any non-negligible modifications to gravity.

It has recently been realised, however, that these experiments have all been performed in the neighbourhood of our solar system and hence they do not rule out modifications to gravity where fifth forces are active over large distances but are suppressed on small scales so that these local constraints are satisfied. Theories such as this are said to possess \textit{screening mechanisms} and regions where these mechanisms act to suppress the force are referred to as being \textit{screened}. Such a mechanism was proposed by Khoury and Weltman in 2004 \cite{Khoury:2003aq} in the form of a scalar-tensor theory with a specific form of the scalar-gravity coupling. This chameleon mechanism, which is also present in $F(R)$ theories \cite{Brax:2008hh}, includes a coupling of the scalar to matter resulting in its properties becoming dependent on the ambient density. Ultimately, this has the effect of screening the force in high density regions of space-time (such as our solar system) and recovering the force predicted by general relativity. Since then, a number of different mechanisms based on similar principles have emerged in the context of the symmetron \cite{Hinterbichler:2011ca} and the environmentally dependent dilaton \cite{Brax:2010gi}. There has been an intense amount of work developing these theories and in looking for observational tests. This has proved difficult since the very mechanism that makes the theory viable and evades solar system constraints tends to suppress any interesting observational features.  Nonetheless, these theories can lead to subtle astrophysical effects such as modifications to galactic dynamics \cite{Adams:2008ad,Hui:2009kc}.

In this paper, we present a new, model independent observational effect of modified gravity with screening mechanisms. Dwarf galaxies in voids can be unscreened in theories with values of the parameters not yet ruled out by observation whereas those in clusters are screened by the presence of their companions. It is possible then, that the stars within these galaxies are partially unscreened and so feel a stronger gravitational force in their outer layers. In order to support themselves against this increased gravitational force they must burn their fuel at a faster rate and hence radiate more energy in a given time period. As a result, their luminosity is greater and their lifetime is reduced. These dwarf galaxies are then more luminous than their screened counterparts in clusters due to cumulative effect of this increased luminosity in their constituent stars. We would then expect that one could construct a statistically based observational test of these theories by looking for systematic offsets between dwarf galaxies situated in voids and those in clusters. Recently, the effect of chameleon theories upon the structure and properties of red giant (RG) stars, which can be resolved in other galaxies, has been investigated numerically by \cite{Chang:2010xh}. 

We analytically investigate this effect in order to gain physical insight into the the effects these theories can have on both stars and dwarf galaxies. For this reason, we consider only main sequence (MS) stars since they are far simpler in their structure and dynamical processes than their post MS counterparts, postponing a full numerical investigation for future work \cite{future_work}. These theories can be described in a model independent manner using two parameters and there is currently some debate as to the observational constraints on these parameters depending on whether one demands that the galaxy is self screened or whether it is screened by other galaxies in the local group. Importantly, the effect presented here differs from those used to place these current bounds and so could form the basis for an independent constraint on values of these parameters that may not yet be ruled out observationally. We have hence consider the entire range of parameters that are currently under debate. 

Main sequence stars are more readily screened than larger stars that have evolved off the MS and therefore an analysis including the effects on these stars could be used to probe parameters and order of magnitude (or even more) lower than have currently been tested and this is investigated in a followup work \cite{future_work}.

This paper is organised as follows. In section \ref{sec:mod_grav} we present the model and derive the screening mechanism, elucidating this with common examples for the unfamiliar reader in appendix A. In section \ref{sec:stars} we calculate the effects of modified gravity on stellar structure and calculate the luminosity enhancement alluded to above and estimate the reduction in the main sequence life time. We also present some numerical calculations in order to reinforce these results although we shall not use them further in this work. We then go on to embed these stars in dwarf galaxies in section \ref{sec:galaxies} and estimate the galactic enhancement in unscreened galaxies due to this increased luminosity and shorter lifetime. We discuss our results in section \ref{sec:conclusion} and speculate on other effects related to these phenomena that may provide additional tests of these theories. Throughout this work we use units such that $\hbar=c=1$ and use the metric signature $(-,+,+,+)$. The reduced Planck mass is $\mpl^2=1/8\pi G$.
 
\section{Modified Gravity with Screening Mechanisms}\label{sec:mod_grav}

\subsection{The Model}

Theories of modified gravity where the extra degrees of freedom are ``screened'' on small scales may arise through the scalar-tensor action 
\begin{eqnarray} \label{eq:field_ac}
S&=\int\mathrm{d}^4 x\sqrt{-g}\left[\frac{\mpl^2}{2}R-\frac{1}{2}\partial_\mu\phi\partial^\mu\phi-V\pp\right]\\
&+S_{\mathrm{m}}\left[\Psi_{i},A^2\!\left(\phi\right)g_{\mu\nu}\right]\,,\nonumber 
\end{eqnarray}
where $\phi$ is a scalar field with self-interactions given by the potential $V(\phi)$. Here, $S_{\rm m}$ is the matter action, $\Psi_{i}$ are the matter fields, which are minimally coupled to the conformally rescaled metric $\tilde{g}_{\mu \nu} \equiv A^2(\phi)g_{\mu\nu}$. The metric $g_{\mu\nu}$ is known as the Einstein frame metric whilst $\tilde{g}_{\mu \nu}$ is referred to as the Jordan frame metric\footnote{Our convention differs from that of \cite{Hui:2009kc} (HNS) by $g_{\mu\nu}^{\rm HNS} =g_{\mu\nu}$, $A^2(\phi) = \Omega(\varphi)^{-2}$ and $M_p\varphi = \phi$.}. Quantities (such as the covariant derivatives, energy-momentum tensor etc.) associated with each metric are distinguished by the use of tildes, with the indices raised and lowered using their respective metrics. The function $A$ is known as the \emph{coupling function} and is assumed to be a weak function of $\phi$, i.e. $A(\phi)\approx 1 +{\cal O}(\phi/\mpl)$ so that metric perturbations in each frame are small.

Matter fields couple minimally to the Jordan frame metric and hence it is the energy-momentum tensor in this frame, $\tilde{T}_m^{\mu \nu} \equiv -2(-\tilde{g})^{-1/2}\delta S_{\rm m}/\delta \tilde{g}_{\mu \nu}$, which is conserved $\tilde{\nabla}_{\mu}\tilde{T}_{\rm m}^{\mu\nu}=0$.

In the Einstein frame the energy-momentum tensor is given by
\begin{equation}
T_{\rm m}^{\mu \nu} \equiv -\frac{2}{\sqrt{-g}}\frac{\delta S_{\rm m}}{\delta g_{\mu \nu}} = A^6(\phi)  \tilde{T}^{\mu\nu}_{\rm m}.
\end{equation}
Since $T_{\rm m}^{\mu\nu}$ is explicitly sourced by the scalar field, the energy-momentum tensor in the Einstein frame is not conserved 
\be\nabla_\mu T_{\rm m}^{\mu\nu}= \frac{\dd \ln A(\phi)}{\dd \phi} T_{\rm m}g^{\mu\nu}\partial_\mu\phi,\ee
where $T_{\rm m}$ denotes the trace (with respect to $g_{\mu\nu}$) of the energy-momentum tensor.
The relation between the Einstein and Jordan frame energy-momentum tensor is 
\begin{equation}
{T_{\rm m}}^{\;\mu}_{\;\nu} = A^4(\phi)\tilde{g}_{\alpha\nu}\tilde{T}^{\mu\alpha}.
\end{equation}

In this paper, we will work exclusively in the Einstein frame. In this frame there is an explicit coupling of the fifth force carrier $\phi$ to the matter tensor in its equation of motion \footnote{The presence of the trace of the Einstein frame energy-momentum tensor $T_m = g_{\mu\nu}T^{\mu\nu}_m$ comes from a functional chain rule when we vary the action with respect with $\phi$. It is clear since radiation is conformally coupled, $T_m^{rad} =0$ and hence do not enter into the equation of motion of $\phi$.} 
\begin{align}
\Box \phi &= \frac{\dd V}{\dd\phi}\pp - \frac{\dd A}{\dd \phi} T_{\rm m}\nonumber \\
 &\equiv \frac{\dd V_{\rm eff}}{\dd \phi}\left(\phi;-T_{\rm m}\right)\,.\label{eq:field_mot1}
\end{align}
Meanwhile, varying the action eqn. (\ref{eq:field_ac}) with respect $g_{\mu\nu}$ gives the modified Einstein equation
\begin{equation}
G^{\mu}_{\,\,\,\nu} =R^{\mu}_{\,\,\,\nu}-\frac{1}{2}R \delta^{\mu}{}_{\nu} = \mpl^2\left[T_m^{\mu}{}_{\nu} + {T_{\phi}}^{\mu}{}_{\nu}\right], 
\end{equation}
where the scalar energy-momentum tensor is
\begin{equation}
T_{\phi}^{\mu}{}_{\nu} = \nabla^{\mu}\phi \nabla_{\nu}\phi - \delta^{\mu}{}_{\nu} \left[\frac{1}{2}(\nabla \phi)^2 + V\pp\right].
\end{equation}

From this, it is clear that the scalar field $\phi$ experiences an effective potential $V_{{\rm eff}}(\phi;-T_m)$ which depends not only on $\phi$ but also the trace of the matter energy-momentum tensor $T_{\rm m}$. In this work, we will assume that the matter is non-relativistic dust as in most astrophysical systems. The usual practice is to define the matter density via $T_m = -\rho$, however, as we have seen above, this quantity is not conserved in the Einstein frame. Now if matter is non-relativistic in the Jordan frame, $\tilde{\rho}=-\tilde{g}_{\mu\nu}\tilde{T}_m^{\mu\nu}$, then it can be shown that the quantity ${\cal T}_m^{\mu\nu} \equiv  A^5(\phi)\tilde{T}_m^{\mu\nu}$ is covariantly conserved in the Einstein frame provided that it is also non-relativistic in that frame\footnote{Such ``conservation laws'' are not true in general, for example, if the matter is slightly relativistic or when observers are highly boosted.}. If one then defines the energy density in the Jordan frame, $\tilde{\rho}=-\tilde{T}_m$, which is covariantly conserved in this frame, then one can show\cite{Waterhouse:2006wv} that the 'density' associated with ${\cal T}_m=-A^3(\phi)\tilde{\rho}$ is (non-relativistically) conserved in the Einstein frame. For this reason, we shall work with this quantity as the density and not the one defined using $T_m$. From here on we shall use the symbol $\rho$ to denote this quantity and refer to it as the matter density. Since, to leading order, $A(\phi)\approx 1$ these two quantities do not differ from each other significantly, although one should bear in mind the discussion above. With this definition, the effective potential is
\begin{equation}
 V_{\rm eff}(\phi;\rho)=V(\phi)+\rho A(\phi).\label{eq:eff_pot}
\end{equation}

The geodesic equation for a test particle made up of $\Psi_i$ in the Einstein Frame then gives us the modified force law in the non-relativistic weak field and static limit\cite{Waterhouse:2006wv}
\begin{equation}
|F_{\phi}| = \frac{\beta(\phi)}{\mpl}\nabla \phi~,\label{eq:forcescalar}
\end{equation}
where the $\phi$ field explicitly appears as a source term as expected. The quantity
\begin{equation}
 \beta(\phi)=\mpl\frac{\dd A(\phi)}{\dd \phi} \label{eq:beta_def}
\end{equation}
then characterises the strength of this force and should only be weakly dependent on $\phi$ if $A(\phi)$ is not to differ too greatly from $1$. 

\subsection{The Screening Mechanism} \label{sec:screen_mech}

In this section, we review the physics behind the screening mechanism and derive an estimate for deciding whether an object is screened or not.

The qualitative features of the screening mechanism are as follows: Suppose that the effective potential \ref{eq:eff_pot} possesses a minimum. The density dependence of the potential has the effect that this minimum occurs at different values of $\phi$ in regions of differing densities. If the field value in high densities is such that the additional force (\ref{eq:forcescalar}) is negligible then any fifth forces will be screened. The form of eqn. (\ref{eq:forcescalar}) reveals that there are two possible methods of accomplishing this. Firstly, the mass of small oscillations about the minimum is proportional to the density. If the density is high enough such that this mass is large compared with the inverse of the length scale associated with the high density region then the field will only vary from its minimum by a small amount over the entire region and so the fieled gradient will be negligible. This can equivalently be thought of as a Yukawa suppression. This is the mechanism employed by chameleon theories\cite{Khoury:2003aq}. Secondly, if $\beta(\phi)$ is close to zero in high density regions then the additional force will again be negligible. This is the mechanism employed by the symmetron \cite{Hinterbichler:2010es} and the environmentally dependent dilaton \cite{Brax:2010gi}.In principle, one could construct a theory where both of these effects are present but, to date, only theories where one of the effects are at play have been considered. Of course none of these mechanisms can operate if the field cannot reach the value which minimises the effective potential and so only regions where this is possible can effectively screen the fifth force. We thus require the field to minimise its effective potential throughout the interior of any high density region if the screening is to be effecient; this is illistrated in fig. (\ref{fig:screening}). In appendix A we provide a brief overview of each of these mechanisms for the unfamiliar reader.

\begin{figure}
\label{fig:screening}
\subfigure[Screened]{\label{fig:screened}\includegraphics[width=0.5\textwidth]{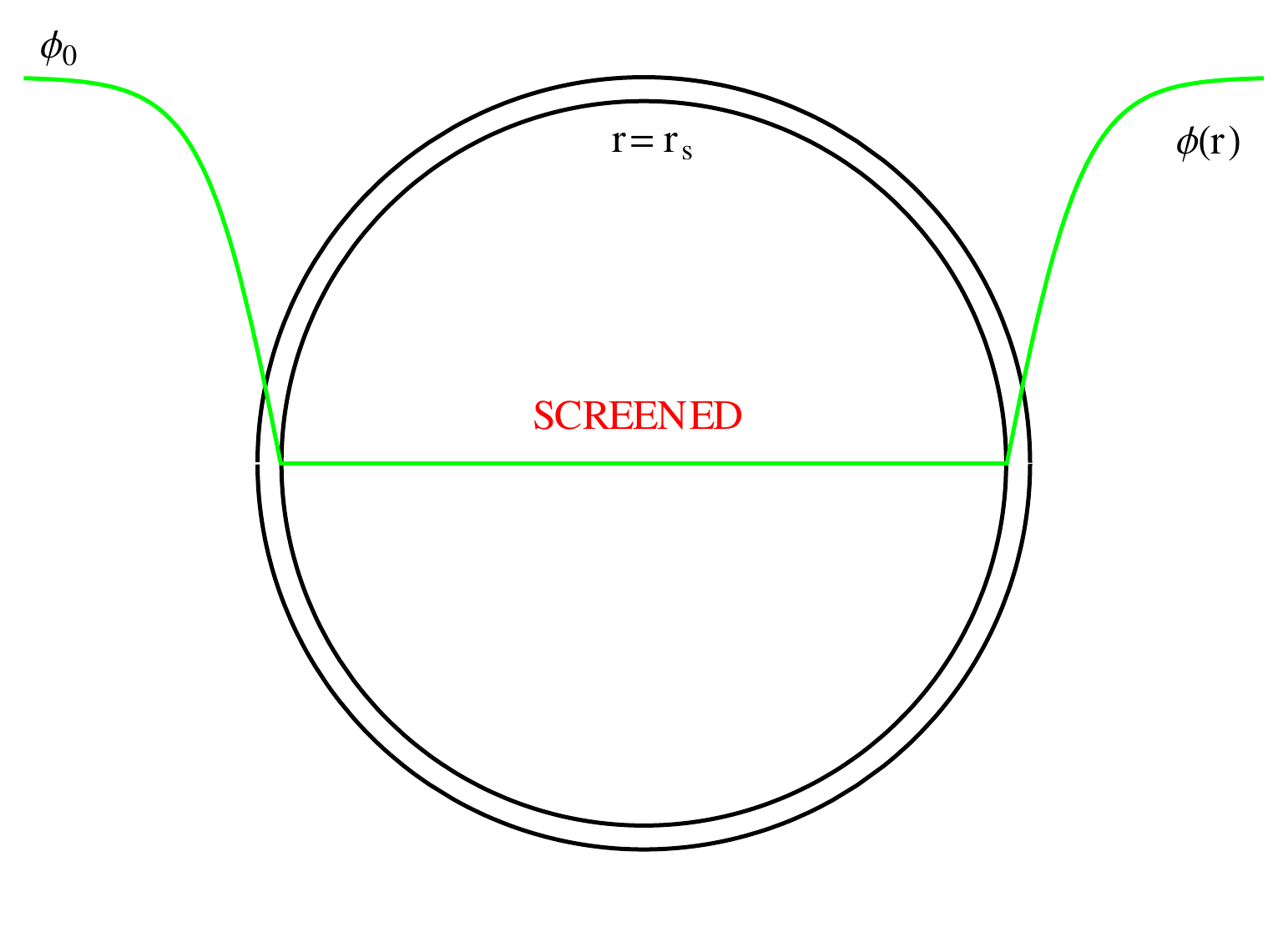}}
\subfigure[Unscreened]{\label{fig:unscreened}\includegraphics[width=0.5\textwidth]{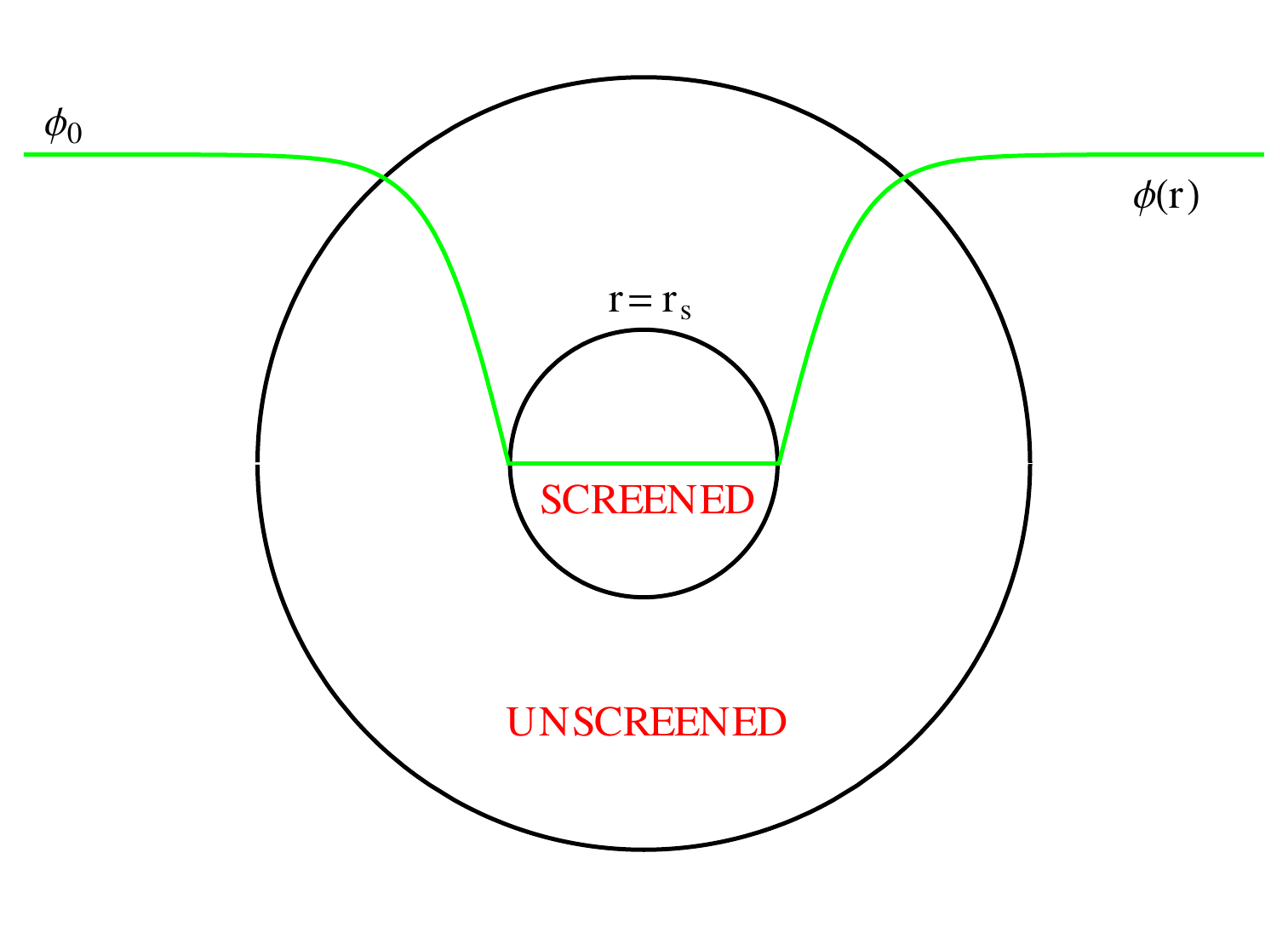}}
\caption{The field profile inside a spherical overdensity embedded in a homogeneous background. The field minimises its effective potential at a distance $r_{\rm s}$ from the centre. Far from the overdensity, the field asemptotically aproaches the value which minimises the effective potential in the background, $\phi_0$. The region $r>r_{\rm s}$ is known as the \emph{unscreened} region and here the fifth force is $\mathcal{O}(1)$. In the region interior to $r_{\rm s}$ the fifth force is negligible and this region is said to be $\emph{screened}$.}
\end{figure}

With the above considerations in mind, we proceed to elucidate the screening mechanism in detail. We consider a homogeneous cosmological background with density $\rho_0$ and assume that sufficient time has passed such that the scalar field has relaxed to the minimum of the effective potential $\phi_0$ in such a background. The scalar in this case possess no dynamics, and hence carries no fifth force while its potential $V_{\rm eff}(\phi_0)$ can act as a cosmological constant term which is small. We will use the subscript $0$ to define quantities in the cosmological background, i.e. $\beta_{0}=\beta(\phi_0)$ and $m_0 \equiv d^2V(\phi_0)/d\phi^2 $ from now on.  

Next, consider a small spherically symmetric overdensity perturbation $\delta \rho(r)$ of radius $R$ in the matter density around this background, which induces a perturbation in the field
\begin{equation}
\phi = \phi_0 + \delta \phi. \label{eq:pertscalar}
\end{equation}
The field inside the overdensity then attempts to relax to its new minimum and the resulting field gradient generates a fifth force $f_{\phi}$ on a test particle (see eqn. (\ref{eq:forcescalar})). If the perturbation is sufficiently large however, the field would reach its new minimum inside the perturbation resulting in a constant radial profile towards the centre and hence no extra fifth force is felt in this region. This central idea behind the screening mechanism: if the perturbation is large enough that the field can reach its new minimum (where, as discussed above, the fifth force is negligible) efficiently then there is only a small field gradient due to the interpolation between the two minima. If, on the other hand, the perturbation is small then the field will not differ significantly from its value in the background and the resulting field gradients will give rise to appreciable fifth forces. Let us investigate under which the perturbation is large enough to screen any fifth forces.

The same over-density $\delta \rho$ sources a perturbation in the Newtonian potential, $\Phi_{\rm N}$, which produces a gravitational force
\be 
\vec{F}_{\rm grav} = \vec{\nabla}\Phi_{\rm N}. \label{eq:forcegrav}
\ee

The Newtonian potential, in turn, obeys the Poisson equation
\begin{equation}
\vec{\nabla}^2 \Phi_N = 4\pi G \delta \rho(r) + ... \label{eq:poisson}
\end{equation}
where we have added the ellipses $...={\cal O}(\mpl^{-2} (\nabla \phi)^2) + \frac{1}{2}V(\phi)$ to remind the reader that, although in the Einstein frame the scalar field does source the Newtonian potential, we neglect them since the gradient terms are second order (recall that we have assume that $\phi$ is small) and its potential is small by construction. \footnote{The innocuous nature of this equation tends to hide some important subtleties. Firstly, the Poisson equation in the Jordan frame possesses similar additional terms which can be neglected. In other words, the Poisson equation is approximately identical in both frames. This is a consequence of our imposition that $A(\phi) \approx 1$. Whilst in general this approximation does not have to be true for the screening mechanism to work, in practice its imposition allows us to treat non-relativistic matter fields in a similar way in both frames. In other words, dust behaves more or less like dust in both frames etc. although they are, of course, sourced differently by the energy density defined in their respective frames. Secondly, as an aside, note that second order terms with time derivatives do not appear in this equation due to our assumption that the system is static.}

In the static limit, where the dynamical time scale of the perturbation is far greater than the time scale on which the field relaxes to its minimum (this will always be the case in this work), the equation of motion, eqn. (\ref{eq:field_mot1}), for $\phi$ is $\nabla^2 \phi = dV_{\rm eff}/d\phi$. Inserting the perturbation eqn. (\ref{eq:pertscalar}) into this equation gives
\begin{eqnarray} 
\vec{\nabla}^2 (\phi_0+\delta \phi)   = \frac{dV}{d\phi}(\phi_0+ \delta \phi)  + \frac{\beta(\phi_0+\delta \phi)}{\mpl} (\rho_0+\delta \rho). \label{eq:phi:2}
\end{eqnarray}
There are now two limits depending on the amplitude of the perturbation. In the presence of a small perturbation, we intuitively expect the field perturbation to be small $|\delta \phi| \ll \phi$. We can then linearise Eq. (\ref{eq:phi:2}) to give:
\begin{equation}
\vec{\nabla}^2 \delta \phi\approx  m^2_{0}\delta\phi + \frac{\beta_0}{\mpl} \delta \rho(r)
\end{equation}
where we have dropped terms proportional to $\dd\beta/\dd \phi$ as we have argued earlier that $\beta$ is only weakly dependent on $\phi$.  Furthermore on scales longer than the Compton wavelength of the scalar  $R \ll 1/m_0$,  the effective mass term is negligible leaving us with 
\begin{equation}
\vec{\nabla}^2 \delta \phi \approx \frac{\beta_0}{\mpl}\delta \rho(r). \label{eq:scalarsourceunscreened}
\end{equation}

In the other limit when the perturbation $\delta \rho$ is large, the field will be able to attain it's new minimum and so the perturbation is also large. In typical theories of interest the field value at the minimum in high density regions is much less than that low densities and so $|\delta \phi| \approx \phi_0$. At distances deep enough in the interior of the perturbation such that the field reaches its new minimum the derivative of the effective potential vanishes and (see for example \cite{Hui:2009kc})
\begin{equation}
\frac{dV}{d\phi}\approx -\frac{\beta(\phi)}{\mpl}\rho(r)
\end{equation}
so that
\begin{equation}
\vec{\nabla}^2 \delta \phi \approx 0
\end{equation}
i.e. the field gets trapped inside the new minimum at some \emph{screening radius} $r<r_{\rm s}$. The field profile becomes constant if $r$ is decreased further leading to no fifth forces. We say that the region inside a sphere of radius $r_s$ is \emph{screened}. In general, $r_{\rm s}$ can be zero (\emph{fully screened}), $0<r_{\rm s}<R$ (\emph{partially screened}) or undefined (\emph{unscreened}). Unscreened objects will feel the full effect of the modified gravity, while fully screened objects are blind to its existence.

Combining these two limits gives an approximation for $\delta \phi$ valid for $r \ll m^{-1}_0$:

\begin{equation}
  \vec{\nabla}^2  \delta \phi \approx \left\{
  \begin{array}{l l}
    \beta_0 \delta \rho(r)/\mpl & \quad r_{\rm s} < r \ll m^{-1}_0\\
    0 & \quad r < r_{\rm s}\\
  \end{array}\right. , \label{eq:scalarpoisson}
\end{equation}

By noting that the density perturbation $\delta \rho$ is related to the Newtonian potential via the Poisson equation eqn. (\ref{eq:poisson}), this equation for $\delta \phi$ can easily be integrated to find the field profile in terms of the Newtonian potential $\Phi_{\rm N}$. For $r_{\rm s}>0$, this is
\begin{widetext}
\begin{equation}
\delta \phi(r) \approx -\phi_0 + 2\beta_0 \mpl \left[\Phi_{\rm N}(r)-\Phi_{\rm N}(r_{\rm s}) + r_{\rm s}^2\Phi^{\prime}_{\rm N}(r_{\rm s})\left( \frac{1}{r}-\frac{1}{r_{\rm s}}\right) \right]H(r-r_{\rm s}), \label{eq:phiform1}
\end{equation}
\end{widetext}
where $H(x>0) = 1$ and $H(x<0) =0$ is the Heaviside function. Taking the limit of the above as $r/r_s \rightarrow \infty$ (and hence $\beta_0 \mpl \Phi_N(r) = \delta\phi(r) \rightarrow 0$) we can find an expression for the screening radius, which is defined implicitly by a function of the constants and the Newtonian potential,
\begin{equation}
\chi_0\equiv\frac{\phi_0}{2\beta_0 \mpl} = -\Phi_{\rm N}(r_{\rm s}) - r_{\rm s}\Phi_{\rm N}^{\prime}(r_{\rm s})\geq0 \label{eq:screenradius}.
\end{equation}
In principle, we can use the Poisson equation eqn. (\ref{eq:poisson}) in conjunction with some density profile $\rho$ to solve for $r_{\rm s}$. Notice however, that since $\Phi_N'>0$, $r_{\rm s}>0$ and $\Phi_N <0$, it is clear that there exist no solutions (i.e. $r_s<0$) when 
\begin{equation}
|\Phi_N(r_s\rightarrow 0)| < \frac{\phi_0}{2\beta_0 M_p} \label{eq:chi0def}.
\end{equation}
In this case, the object is fully unscreened. It is important to note that $\chi_0$ is completely defined by the parameters of the theory in question. It follows then that if the surface Newtonian potential $\Phi_N > \chi_0$, the screening radius $r_s>0$ and the object will become gradually screened. This result gives us a rough guide on whether an object is fully screened or not, simply estimate the depth of the gravitational potential at the surface of this object (including the mass from the surroundings and its own) and see if eqn. (\ref{eq:chi0def}) is satisfied. Note that if $r_{\rm s}=R$ then the fifth force is completely suppressed and $\vec{F}_{\phi} = 0$ everywhere.\newline
Now that we have the complete field profile we can calculate the fifth force explicitly. For spherically symmetric bodies we have \be \frac{\dd \Phi_{\rm N}}{\dd r} = \frac{G M(r)}{r^2},\ee where $M(r)$ is the mass enclosed within a shell of radius $r$, and so differentiating eqn. (\ref{eq:phiform1}) we find the total force per unit mass (gravity and modification) in the unscreened region is
\begin{align}\begin{split}\label{eq:force}
F &= F_{\rm N}+F_{\phi}\\ &= \frac{\dd \Phi_{\rm N}}{\dd r} +\frac{\beta_0}{\mpl}\frac{\dd \phi}{\dd r}\\ & = \frac{GM(r)}{r^2}\left[1+\alpha_0\left(1-\frac{M(r_{\rm s})}{M(r)}\right)\right],
\end{split}\end{align}
where $\alpha(\phi)\equiv 2\beta^2(\phi)$, which defines and effective gravitational force \be\label{eq:g(r)} F=\frac{G(r)M(r)}{r^2}. \ee  The effect of modified gravity is then to produce an effective change in the gravitational constant $G\rightarrow G(r)\equiv G(1+\alpha_{\rm eff}(r))$ in the unscreened region, where
\be\label{eq:aleff}
\alpha_{\rm eff}(r) = \alpha_0\left(1-\frac{M(r_{\rm s})}{M(r)}\right).
\ee One can think of the quantity $1-M(r_s)/M(r)$ as a scalar charge, the existence of which is the source of equivalence principle violation in such theories. We can see that completely unscreened objects will feel a fifth force that is a factor $\alpha_0$ that of the Newtonian force. We may then describe all theories with screening mechanisms in a model independent manner using two parameters: $\alpha_0$ and $\chi_0$. $\alpha_0$ specifies the strength of the fifth force while$\chi_0$ (known as $f_{R0}$ in the context of $F(R)$ models) controls the extent of the screening and can take on a range of different values in any one specific class of models, depending on both the position of the minimum and the coupling. Currently, there is some debate on the value of $\chi_0$ that is consistent with local observations. If one demands that the solar system is screened by the potential of the local group galaxies then the lowest bounds come from cluster abundance statistics \cite{Schmidt:2008tn}, which yield $\chi_0\sim\mathcal{O}($10$^{-4})$. If, on the other hand, one demands that the sun is self-screening then, as explained above, we must take $\chi_0\sim\Phi_{\rm N}(R_\odot)\sim\mathcal{O}($10$^{-6})$ \cite{Khoury:2003rn}.

\section{The Effect of Modified Gravity Upon Stellar Structure}\label{sec:stars}

In this section we describe how these theories with screening mechanisms alter the structure of stars compared with standard gravity. Before examining the details, it is instructive to first consider the effects in a heuristic manner. If the star is self-screening then we expect any effects to be negligible (or at least unobservable). If, on the other hand, the star is at least partially unscreened then in some part of its outer layers, the gravity will be stronger than the standard case. Stars support themselves against gravitational collapse by hydrogen fusion in their core to release energy and provide an outward pressure. We would then expect these reactions to proceed at a faster rate in order to combat the increased gravity. This will have two effects. Firstly, the star is more luminous and secondly, it will deplete its fuel supply at a faster rate and will hence have a shorter lifetime.  

This effect was first noted by previous work by Chang and Hui \cite{Chang:2010xh} who found that solar mass $M=M_{\odot}$ stars in the Red Giant Branch (RGB) phase of their life have a large increase in its stellar envelope reduces the depth of its self gravity potential, resulting in its outer ``mantle'' layers becoming unscreened. In this section, we show that \emph{even in stars on the main sequence, a substantial part of the star's interior can be unscreened}. Indeed, we will show that $r_s/R_{\odot} \approx 0.3$ for in solar mass stars if $\chi_0\approx 10^{-6}$. We emphasise that these are ${\cal O}(1)$ effects, however, they are also highly degenerate with other stellar properties such as metallicity and other environmental effects. In this paper, we will argue that while it is hard to disentangle these effects in individual star systems, the gross effects of such modified gravity stars on galaxies may be observable. 

\subsection{The Theory Stellar Structure in Modified Gravity}\label{sec:over_star}

The standard approach to stellar structure is to consider static, spherically symmetric stars of mass $M$ and radius $R$. In order for the star to remain static, the outward force due to the pressure generated by the nuclear reactions must balance the gravitational force $F(r)$ (per unit mass). This condition is expressed through the hydrostatic equilibrium equation (HSE)
\be\label{eq:hydrostatic} \frac{\dd P}{\dd r}=-F(r)\rho(r) \ee
where $\rho(r)$ is the density at a distance $r$ from the centre. The radius $R$ is defined to be the value of $r$ at which the pressure becomes zero, although this can be ambiguous and in more realistic models a definition involving the optical depth is more appropriate. The mass $M(r)$ enclosed by a shell of radius $r$ (so that $M= m(R)$) may be found by integrating the density over the volume enclosed by $r$, which leads to the continuity equation,
\be
\frac{\dd M}{\dd r} = 4\pi r^2\rho.\label{eq:continuity}
\ee
By considering photon propagation in the interior of the star, we obtain the radiative transfer equation,
\be\label{eq:radiative}
\frac{\dd T}{\dd r} = -\frac{3}{4 a} \frac{\kappa(r)}{T^3} \frac{\rho(r) L(r)}{4\pi r^2},
\ee
where $L(r)$ and $T(r)$ are the luminosity and temperature respectively at coordinate $r$. The quantity $\kappa$ is known as the opacity and represents the cross section per unit mass for radiation absorption; it is generally a function of the temperature and density. Finally, if energy is generated at a rate $q(r)$ per unit volume by nuclear (or possibly other) processes then the luminosity gradient is determined by the energy generation equation,
\be
\frac{\dd L}{\dd r} = 4\pi r^2 \rho q(r).\label{eq:engen}
\ee
These are the equations of stellar structure, which describe the relationships between the various thermodynamic and physical quantities which are required in order for the star to remain in a static configuration. As it stands, these equations do not close and so we require an equation of state for both the opacity and the pressure. We shall specify these later when we discuss scaling solutions and the Eddington standard model.%

Note that since Newton's constant, $G$, only appears in the HSE equation, the presence of modified gravity only changes that particular equation and not the others \footnote{One can also imagine non-trivial couplings of the scalar to other degrees of freedom such as $\phi F_{\mu\nu}F^{\mu\nu}$ where $F_{\mu\nu}$ is the electromagnetic field strength tensor, which would then change the other equations. We will not consider such couplings in this work although such couplings do indeed exist \cite{Brax:2010uq}.}. In other words, the total force acting on the star is the gravitational force and the fifth force given by eqn. (\ref{eq:force}). Eqn. (\ref{eq:hydrostatic}) then requires that
\begin{equation}
\frac{\dd P(r)}{\dd r} = -\left[\frac{GM(r)}{r^2}\left[1+\alpha_0\left(1-\frac{M(r_{\rm s})}{M(r)}\right)\right]\right] \rho(r). \label{eq:modhydrostatic}
\end{equation}
To find the gravitational force, we integrate the Poisson equation eqn. (\ref{eq:poisson}) and use the continuity equation eqn. (\ref{eq:continuity}) to obtain the solution to the gravitational potential as a function of mass $M(r)$,
\be \label{eq:stellarstructure}
\frac{\dd \Phi_{\rm N}}{\dd r} = \frac{G M(r)}{r^2}, \qquad M(r) = 4\pi \int_{0}^{r} r^{\prime\,2}\dd r^{\prime}\, \rho(r^{\prime}). \label{eq:stellarGravity}
\ee
Finally, in order to close the system of equations, we find $r_{\rm s}$ from Eq. (\ref{eq:screenradius}), which after integrating by parts and using the Poisson eqn. (\ref{eq:poisson}) yields an implicit solution for $r_{\rm s}$
\be
4\pi G \int_{r_{\rm s}}^{R} r\rho(r) \dd r = \chi_0 \equiv \frac{\phi_0}{2\beta_0 \mpl}. \label{eq:mesacon}
\ee
The implicity of the solution for $r_{\rm s}$ means that we have to iterate to find the complete solution. We will do this in the next section for a polytropic star.

\subsection{Scaling Relations}

In the next section we shall derive the structure of partially screened stars but before doing so we can extract all of the essential physics by considering the simpler case of completely unscreened stars where the effect of modified gravity is to change the value of Newton's constant by a constant factor $G\rightarrow G(1+\alpha_0)$. In this case, we can use dimensional analysis to find the scaling of the stellar luminosity with the gravitational constant at fixed mass. By replacing each quantity $Y_i$ where $Y=\{M(r),P(r),\rho(r),T(r),L(r)\}$ by its characteristic value (say stellar mass or central density) in the equations of stellar structure and replacing the derivatives by ratios (remembering to account for the correct sign depending on whether a quantity increases towards the centre or surface) we can obtain \textit{scaling relations} between these variables. For example, the hydrostatic equilibrium equation eqn. (\ref{eq:hydrostatic}) becomes \be P_{\rm c}\propto \frac{GM\rho_{\rm c}}{R},\ee where $P_{\rm c}$ and $\rho_{\rm c}$ are the central pressure and density respectively. This process may be repeated for the other equations to find the complete set of scaling relations. In order to complete this analysis we need to specify the form of the pressure. We consider two different sources of pressure, namely that due to radiation,
\be\label{eq:prad}
P_{\rm rad} = \frac{1}{3}aT^4
\ee
and that due to the gas, which we take to be ideal:
\be\label{eq:pgas}
P_{\rm gas} = \frac{k_{\rm B}\rho T}{\mu m_{\rm H}},
\ee
where $\mu m_{\rm H}$ is the mean molecular mass, which we take to be constant. Low mass stars are supported mainly by gas pressure and in this case the luminosity scales as $L\propto G^4M^3$. More massive stars are hotter and are therefore supported predominantly by radiation pressure with a mass-luminosity relation $L\propto GM$. Physically, stars supported by radiation pressure absorb radiation in their interior to prove the outward force needed to prevent gravitational collapse and less reaches the surface. If we then consider two stars of equal mass, one screened and the other completely unscreened we find a luminosity enhancement of
\be
\frac{L}{L_{\chi_{0}=0}} = \left\{
  \begin{array}{l l}
   (1+\alpha_0)^4 & \quad \text{Low mass stars}\\
    1+\alpha_0 & \quad \text{High mass stars}\\
  \end{array} \right.
\ee
in unscreened stars.
Clearly for typical values of $\alpha_0$, which is $1/3$ for $F(R)$ gravity \cite{Brax:2008hh}, this increase is of order $1$ and so this effect could constitute a potential observational test of these theories.

\subsection{The Eddington Standard Model and Partially Screened Stars}\label{sec:edd_mod}

Stars on the main sequence are in general supported by a combination of radiation and gas pressures $P(r) =  P_{\rm rad}+P_{\rm gas}$. The set of equations eqn. (\ref{eq:hydrostatic})-(\ref{eq:engen}), combine with a set of equations of state (and modified gravity) represent a highly complicated set of differential equations which are difficult to solve in general. A complete description of stellar structure will require a numerical method, which we will present in the section after this. 

In this section, we will instead  solve these set of equations in a particularly simple approximation known as the Eddington Standard model. While this model does not describe true stars, it is a fair approximation to main sequence stars. More importantly, it captures the essential physics and we can gain a lot of intuition and understanding of the physics underlying such systems under the influence of modified gravity.  

In the Eddington Standard model, we make the assumption that the radiation entropy per unit mass is constant throughout  $S_{\rm rad} = 4a T^3/3\rho =  {\rm const}$. 
This greatly simplifies the equations by decoupling the HSE from both the radiative transfer eqn. (\ref{eq:radiative}) and the energy generation eqn. (\ref{eq:engen}). This assumption constrains $T$ to be a function of $\rho$ alone and therefore reduces the number of independent degrees of freedom in the system. Furthermore, by writing $P_{\rm rad} = (1-b)P$, $P_{\rm gas} = b P$ where $b$ measures the fraction of gas pressure support and assuming that $P_{\rm gas}$ is given by the ideal gas law, the equation of state becomes polytropic i.e. $P=K\rho^{4/3}$ \footnote{This is a polytrope of index $n=3$ which usually is a very good description of a radiation supported star. In addition, standard stellar structure notation uses $\beta$ for fractional gas pressure support. Here, we use $b$ instead in order not to confuse it with the modified gravity force coupling function. Other polytropic indices have been considered in the literature and our method and results generalise in a straight forward manner to these cases, however, for simplicity we present only the specialised case of $n=3$}, as follows 
\begin{eqnarray}
P& =& P_{\rm rad}+P_{\rm gas} = \frac{a}{3}T^4 + \frac{k_{\rm B}}{\mu m_{\rm H}}\rho T\nonumber \\ &=& K(b) \rho^{4/3}, \\
K(b) &=& \left[ \frac{3}{a}\left(\frac{k}{\mu m_{\rm H}}\right)^4 \frac{(1-b)}{b^{4}}\right]^{1/3}.
\end{eqnarray}
Hence we simply need to solve the HSE to obtain $P(r)$ and $\rho(r)$ which would then fully describe the star. In addition, the power law dependence of $\rho$ means that the solution, in general without modified gravity, is self-similar (i.e. one can describe using dimensionless coordinates). The rescaled equations in this case is called the Lane-Emden equation.  On the other hand, the presence of a screening radius $r_{\rm s}$ weakly breaks this self-similarity, resulting in a modified form of Lane-Emden equation which we will solve below. Note that while $b$ depends on the properties of the star (and modified gravity), it is a constant for a given star in the Eddington Standard mode, hence the space of Eddington Standard model solutions is spanned by this single family of parameters, $b$.

Once a solution $\rho(r)$ and $P(r)$ is obtained (and hence $T(r)$ by the assumption of constant entropy density), we substitute it back into the radiative transfer eqn. (\ref{eq:radiative}) to obtain the total luminosity at the star surface where $r=R$. Further assuming that the opacity is a constant and given by the electron scattering opacity taking $\kappa_{\rm op}(R) \approx \kappa_{\rm es}$ we obtain the luminosity in terms of $\alpha_{\rm surf}\equiv \alpha_{\rm eff}(R)$ and $M$:
\be 
L = \frac{4\pi c (1-b)(1+\alpha_{\rm surf}) GM}{\kappa_{\rm es}}, \label{eq:modlum}\ee
where
\be
\alpha_{\rm surf} &= \alpha_0\left[1-\frac{M(r_{\rm s})}{M}\right]=\alpha_{\rm eff}(M). 
\ee

Before proceeding to solve the Lane-Emden equation, we pause to discuss the nature of these approximations. The decoupling of the energy generation and radiative transfer equations from the mechanical structure equations have allowed us to close the equations. This means that we do not include any information about the nuclear processes occurring in the core, nor do we account for any chemical composition or gradient which may change over time. We have also used a constant opacity, which in general has a temperature and density dependence and have therefore ignored the effects of bound-bound and bound-free transitions, as well as H$^{-}$ effects in the outer layers. 

Finally, we have not included the effects of convection, which tends to be important in the outer layers of stars. Nonetheless, the Lane-Emden solution is a good approximation to stars near the zero age main sequence (ZAMS). For this reason, we will not include the radial dependence of the fifth force in the following, but will model it by scaling Newton's constant by a constant factor, $G\rightarrow G(1+\alpha_0)$. When we come to the full numerical problem in later sections we will include the full radial dependence as well as a complete theory of chemical evolution and convection and so will be able to make far more realistic predictions for any evolutionary stage in the stars life.

Let us now solve the equations. What follows is a technical derivation, which the reader may not be interested in and is not required to understand the rest of the paper. Those who are not can safely skip to the next section. 

Since polytropic solutions are spanned by $b$, our goal is to solve for $b$ given $\alpha_0$ and the stellar mass $M$. The Lane-Emden equation can be obtained by first defining a set of dimensionless variables -- define $P$, $\rho$ and $T$  at $r=0$ to be $P_{\rm c}$, $\rho_{\rm c}$ and $T_{\rm c}$ respectively as the central values.  Then we can define a dimensionless radial coordinate $\xi$ by:
\begin{equation}
r = \left(\frac{P_{\rm c}}{\pi G \rho_c^2}\right)^{1/2} \xi \equiv r_{\rm c}\xi, \nonumber
\end{equation}
where $r_{\rm c}$ is the characteristic length-scale of the star and the dimensionless structure function 
\begin{equation}
\theta(\xi) = T(r)/T_{\rm c}
\end{equation} 
hence $P = P_{\rm c}\theta^4(\xi)$ and $\rho = \rho_{\rm c}\theta^3(\xi)$.  Inserting these new variables into the HSE eqn. (\ref{eq:hydrostatic}), ignoring the radial dependent term $M(r)/M(r_{\rm s})$, and differentiating it then gives the \emph{modified Lane-Emden equation} for $\theta(\xi)$:
\begin{equation}
\frac{1}{\xi^2} \frac{ \dd}{\dd \xi}\left[ \xi^2 \frac{\dd \theta}{\dd \xi}\right]=\left\{
\begin{array}{l l}
 -(1+\alpha_0) \theta^{3}(\xi),& \quad   r_{\rm s} < r <R,   \label{eqnLE}\\
 - \theta^{3},&\quad r<r_{\rm s},  
 \end{array}\right.
\end{equation}
where the boundary conditions are the stellar radius $R$ is defined by $P(R)=0$ and therefore $\theta(\xi_R)=0$ where $\xi_R=R/r_0$. Since we have scaled out the central pressure and density in terms of $\theta$ we have $\theta(0)=1$ and any sensible pressure profile should be smooth at the centre and so we require $\dd\theta(0)/\dd \xi=0$. This second condition can also be viewed as a statement that there is no mass inside a shell of zero radius; setting $M(0)=0$ in eqn. (\ref{eq:hydrostatic}) gives precisely this condition. With these two boundary conditions, we can proceed to solve the equation, matching up the solutions at $r=r_s$. 

We can similarly rescale the screening radius and the total radius of the star by the characteristic scale $\xi_{\rm s} \equiv r_{\rm s}/r_{\rm c}$ and $\xi_{R} \equiv R/r_{\rm c}$; $\xi_{R}$ is determined by the requirement that $\theta(\xi_{R}) =0$ i.e. $\rho(R) = 0$. The first derivative of the boundary condition is specified by 
\begin{align}
\omega_{R} &= -\xi_{R}^2 \left.\frac{\dd \theta}{\dd \xi} \right\vert_{\xi = \xi_{R}}.\label{eq:omegas_2}
\end{align}
Integrating this equation once (keeping $r_{\rm s}$ and hence $\xi_{\rm s}$ fixed for the moment) and imposing continuity at $\xi_{\rm s}$ requires us to further specify the matching condition
\begin{align}
\omega_{\rm s} &= -\xi_{\rm s}^2 \left.\frac{\dd \theta}{\dd \xi} \right\vert_{\xi = \xi_{\rm s}}. \label{eq:omegas}
\end{align} 
$r_{\rm s}$ (and hence $\xi_{\rm s}$) generally depend on both the mass and the radius of the star. It will turn out, however, that the Lane-Emden scaling constrains $r_{\rm s}$ to depend on the combination $\chi_0/(GM/R)$. It then follows that to fully integrate the Lane-Emden equation we will need to find this quantity. 

Consider the case of standard GR ($\alpha_0 = 0$), we have $\omega_{R} =  \bar{\omega}_{R} \approx 2.02$.  If on the other hand, the star is fully unscreened ($r_{\rm s}=0$), then $\omega_{\rm s}=0$ and by rescaling $\xi$ we have $\omega_{R} \approx 2.02 (1+\alpha_0)^{-1/2}$. Between these two limits, we can define a function that interpolates between $\alpha_b \in (0,\alpha_0)$ : 
\begin{equation}
1+\alpha_b = \left((1+\alpha_0)\frac{\bar{\omega}_{R}}{\omega_{R}+\alpha_0\omega_{\rm s}}\right)^{2/3}, \label{eq:alphab}
\end{equation}
such that when $r_{\rm s}=R$, $\omega_{R}=\omega_{\rm s} = \bar{\omega}_{1}$ we have $\alpha_{b}=0$, while when $r_{\rm s}=0$ the star is fully unscreened and $\alpha_{b}=\alpha_0$. The quantity $\alpha_b$ is the analog of $\alpha_{\rm eff}$ in Lane-Emden rescaled variables and is a function of $b$ hence the its subscript.

We can then find the mass of the star in terms of these variables:
\begin{eqnarray}
M &=& 4\pi \int_{0}^{R} r^2 \rho(r)\dd r = 4\pi \rho_{\rm c} r_{0}^3 \int_{0}^{\xi_{1}} \xi^2 \theta^{3}(\xi)\dd \xi, \nonumber \\
&=& 4\pi K(b)^{\frac{3}{2}} \left(\frac{1}{\pi G}\right)^{3/2} \left[ \frac{\omega_{R} + \alpha_0 \omega_{\rm s}}{1+\alpha_0}\right].  \label{eq:Mass}
\end{eqnarray}
Inverting this gives $b$ as a function of $M$ through a modified form of Eddington's quartic equation
\begin{equation}
\frac{1-b}{b^4} = (1+\alpha_{b})^3 \left(\frac{M}{M_{\rm edd}}\right)^2, \label{eq:ed_quart} 
\end{equation}
with the Eddington mass 
\begin{equation}
M_{\rm edd} = \frac{4 \pi^{1/2}}{G^{3/2}\bar{\omega}_R} \left(\frac{3}{a}\right)^{1/2} \left(\frac{k_{\rm B}}{\mu m_{\rm H}}\right)^{2}, 
\end{equation}
which is defined in terms of fundamental constants and hence  ($M_{\rm edd} = 18.3 \mu^{-2} M_\odot$) is independent of $\alpha_0$. Finally, to close this set of equations we must specify $\xi_{\rm s}$ as a function of the modified gravity parameters $\phi_0$ and $\beta_0$. Using eqn. (\ref{eq:scalarpoisson}) we have
\begin{equation}
 \frac{\beta_0} {\mpl}\vec{\nabla}^2 \phi = \frac{4\pi G \alpha_0 \rho(r)}{\mpl} H(r-r_{\rm s}), 
\end{equation}
from which, using our knowledge that $\phi \rightarrow 0$ as $r \rightarrow \infty$, we find that $\phi_{\rm s} = \phi(r_{\rm s})$ is given by
\begin{eqnarray}
\frac{\beta_0 \phi_{\rm s}}{\mpl} &\approx& \frac{4\pi G r_{\rm 0}^2 \rho_{\rm c} }{\xi_{R}} \frac{\alpha_0}{1+\alpha_0} \left[ \theta_{\rm s} + \omega_{R} - \frac{\xi_{R}}{\xi_{\rm s}}\omega_{\rm s}\right], \nonumber \\
&=& \frac{\alpha_0 GM}{R} \left[\frac{\xi_{R} \theta_{\rm s} + \omega_{R} - \frac{\xi_{R}}{\xi_{\rm s}}\omega_{\rm s}}{\omega_{R}+\alpha_0\omega_{\rm s}}\right]. \nonumber
\end{eqnarray}
The scaled screening radius, $\xi_{\rm s}$, is then determined by  $\phi_{\rm s} = -\phi_0$ and
\begin{equation}
\left[\frac{\xi_{R} \theta_{\rm s} + \omega_{R} - \frac{\xi_{R}}{\xi_{\rm s}}\omega_{\rm s}}{\omega_{R}+\alpha_0\omega_{\rm s}}\right] = \frac{\phi_0}{2\beta_0 \mpl} \frac{R}{GM} \equiv \frac{\chi_0}{GM/R}. \label{eq:LEsuprad}
\end{equation}
This is the analog of the implicit solution for $r_{\rm s}$, eqn. (\ref{eq:mesacon}), in Lane-Emden coordinates, and requires a similar iterative method to solve. If eqn. (\ref{eq:LEsuprad}) has no solutions then $r_{\rm s}=\xi_{\rm s}=\omega_{\rm s}=0$ and the fifth force is unsuppressed. This requires
\begin{equation}
\frac{\chi_0}{GM/R} \geq 1 - \left.\frac{\dd \ln \xi}{\dd \theta}\right\vert_{\xi=\xi_{R}} \approx 4.417,
\end{equation}
for all $\alpha_0$. The numerical value $4.417$ is a property of $n=3$ self-similar (i.e. completely screened) polytropes. Thus if $GM / R > 0.226 \chi_0$ the fifth force is at least partially suppressed. Finally, we can then find $\alpha_{\rm surf}$, which gives us the modification to the luminosity:
\begin{equation}
\frac{\alpha_{\rm surf}}{\alpha_0}  = 1- \frac{M(r_{\rm s})}{M} = \left[\frac{\omega_{R}-\omega_{\rm s}}{\omega_{R}+ \alpha_0}\omega_{\rm s}\right].\end{equation}\newline

The procedure to find the luminosity increase is the following: we first solve the Lane-Emden equation numerically for a set of test matching radii $\omega_{\rm s}$. We then use this function $\theta(\xi;\xi_{\rm s})$ in conjunction with eqn. (\ref{eq:omegas}) to iteratively find $\xi_{\rm s}$ using Eqn (\ref{eq:Mass}) and Eqn (\ref{eq:LEsuprad}), and hence $\alpha_b$ and $\alpha_{\rm surf}$. By repeating this for a range of different masses $M$ and values of $\chi_0$ we numerically fit the luminosity as a function of mass and $\chi_0$. Since the screening radius is degenerate in $\chi_0$ and $M/R$, it is convenient to scan through the quantity $X \equiv \chi_0/(4.417 GM/R)$ such that $0 \leq X \leq 1$. The relation between $\alpha_b$ and $\alpha_0$ can then be recast as $\alpha_b(1+\alpha_b) = \alpha_0(1+\alpha_0)f(X;\alpha_0)$ and that $\alpha_{\rm surf}/\alpha_0 = g(X;\alpha_0)$ where the fitting functions $f$ and $g$ take values between 0 and 1. In doing this, we find $f(X;\alpha_0)$ and $g(X;\alpha_0)$ are independent of $\alpha_0$ for $\alpha_0 \sim \mathcal{O}(1)$ and smaller and are respectively well fitted by the following functional forms
\be \label{eq:fits}
f(x) &=&  x^2(1.94+0.79x -2.91x^2+1.18x^3), \nonumber \\
g(x) &=& \sqrt{-\frac{13}{14} +\sqrt{\frac{169}{196}+\frac{20 x}{7}}}. 
\ee

\begin{figure}[ht] 
   \centering
      \includegraphics[width=3.5in]{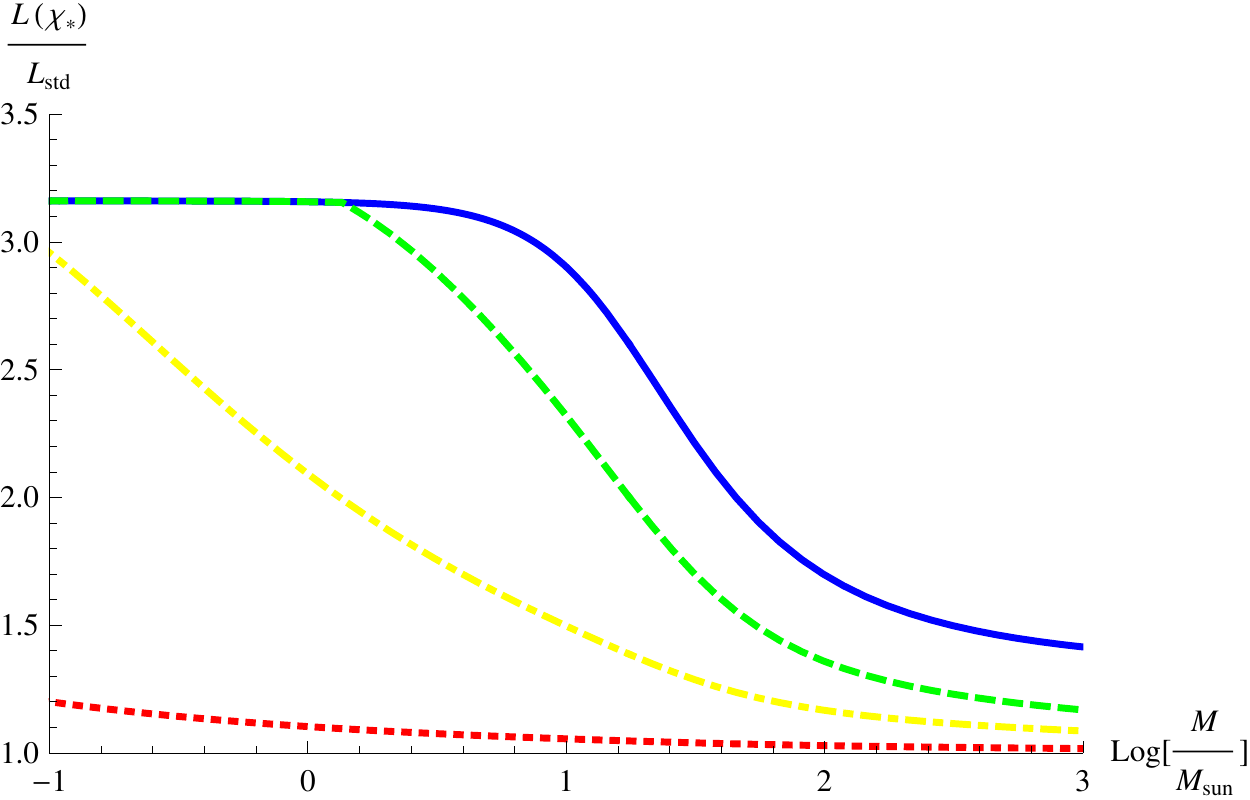}
         \caption{The ratio of the stellar luminosity of a partially screened star compared with one that is completely screened as a function of stellar mass $M$. $L_{\rm std}$ is the luminosity of a completely screened (general relativity) star. From top to bottom: $\chi_0 = (10^{-4},10^{-5},5\times 10^{-6},10^{-6})$.  }
          \label{fig:lumplot}
       \end{figure}

In principle, one could vary $\alpha_0$ to find the luminosity increase as a function of $\alpha_0$ however in this work we consider only $\alpha_0 = 1/3$, corresponding to $F(R)$ gravity and the low background density limit of the environmentally dependent dilaton. The luminosity increase compared to the GR predictions is shown in fig \ref{fig:lumplot}. We can see that the enhancement is much greater in low mass stars in agreement with the simple scaling arguments presented above. Numerically, $(1+\alpha_0)^4=(4/3)^4\approx3.16$ and so we can see that for $\chi_0=$10$^{-5}$ and 10$^{-4}$ the lowest mass stars are entirely unscreened. We also note that at 5x10$^{-6}$ there is an appreciable enhancement but at 10$^{-6}$ we see that there is no significant effect. This is due to the fact that main sequence stars typically have Newtonian potential of order 10$^{-6}$ and so, according the the arguments above, are almost entirely screened. After going off the main sequence, the stars become red giants which have radii of 10 to 100 times that of their main sequence phase and so even if they are screened in the main sequence, they can become unscreened in the red giant phase and show appreciable differences from GR \cite{Chang:2010xh}. 

There is another important consequence of this increased luminosity. Although we have not included the effects of nuclear burning in our model, if the luminosity of a star in modified gravity is higher than the standard case then the extra energy being radiated must come from an increased rate of nuclear fusion in the core. The main sequence lifetime is then shorter. The main sequence lifetime can be well approximated by
\be\label{eq:ms_life}
\tau_{\rm MS} \approx 10\left(\frac{M}{M_{\odot}}\right)\left(\frac{L_{\odot}}{L}\right)\textrm{ Gyr}
\ee
and hence we can see that a fully unscreened star in $F(R)$ gravity will go off the main sequence more than three times as quickly as one in general relativity. Ultimately, as we shall see below, this effect limits the luminosity enhancement in unscreened galaxies since stars go off the main sequence before their enhanced luminosity can contribute significantly. This has important observational consequences since we then expect unscreened galaxies to have hosted many more generations of stars and therefore be more metal enriched than their screened counterparts. 

\subsection{Numerical Simulations of Modified Gravity}

The assumptions of the previous subsection are a good approximation to intermediate main sequence stars. However, as alluded to in the introduction to this section, making further progress and understanding the many degeneracies in stellar systems, requires accurate modelling of the stellar processes and knowledge of the complete evolution of the star from the pre-main sequence until its death. In practice, stellar evolution is determined using numerical evolution codes. In this section, we will take this approach for the case of modified gravity. As we have mentioned, modified gravity only changes the gravitational part of systems of equations, and hence we simply need to modify existing stellar evolution codes to make progress. We have chosen to implement the screening mechanism into the publicly available code MESA-star \cite{Paxton:2010ji}.\newline

MESA is a one dimensional (in that it assumes spherical symmetry) stellar evolution code that numerically calculates the structure and evolution of stars by solving the stellar structure equations in conjunction with a fully consistent implementation of varying opacity laws, nuclear reaction networks, atmospheric models, convective theories, mass loss schemes and rotational dynamics. We refer the interested reader to the instrument paper \cite{Paxton:2010ji}. We will describe our implementation of the screening mechanism into MESA and present some simple results in order to justify the analysis above.\newline

We start by noting that, given a density profile, the screening radius can be found using eqn. (\ref{eq:mesacon}) given a specific value of $\chi_0$. Using this result, we implement the modification of gravity into MESA using the following procedure. MESA divides the star into cells of unequal radius and assigns each cell a set of quantities such as radius, density, opacity etc. and we first integrate the quantity $\rho$ from the surface of the star, cell by cell, until the condition \ref{eq:mesacon} is satisfied. The radius of the cell where this is the case is marked as the screening radius. Next, we update the value of $G$ in each cell using eqn. (\ref{eq:aleff}). The star is then allowed to evolve in time to the next model where this process is repeated. This process is accurate provided that the timescale over which the star evolves is not longer than the timescale on which the effective gravitational constant varies due to a change in the stellar structure. MESA automatically selects a timestep to reduce such numerical errors however it also has controls that allow the user to set the timestep and the number of cells and so an arbitrarily good precision can be achieved.

\begin{figure*}[ht] 
   \centering
      \includegraphics[width=7in]{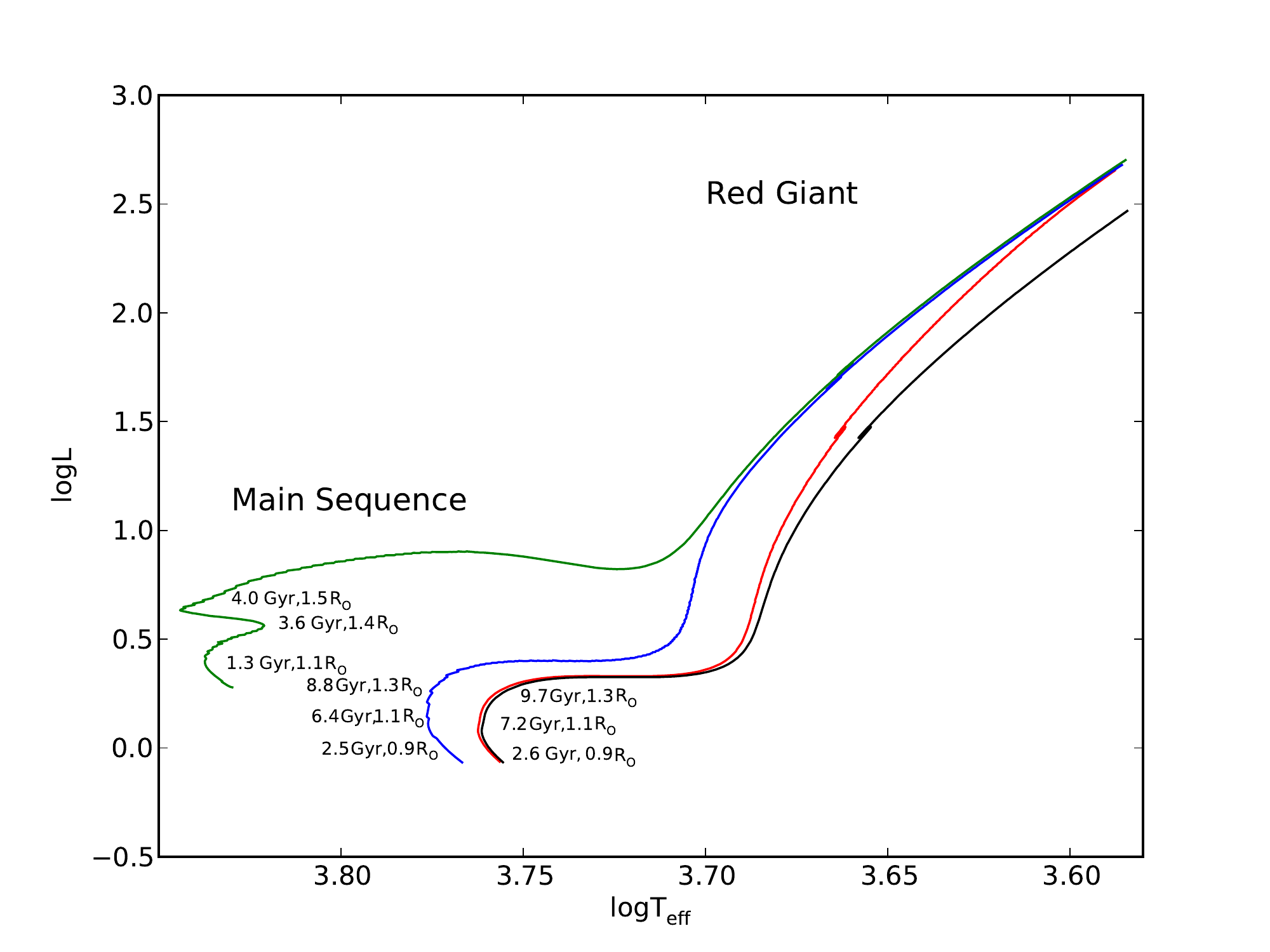}
         \caption{The Hertsprung-Russell diagram for stars of one solar mass with initial metallicity $Z=0.02$. The black line shows the tracks for stars in general relativity while the red, blue and green tracks correspond to stars in modified gravity with $\chi_0 = 10^{-7}, 10^{-6}$ and $5\times10^{-6}$ respectively. The radius and age at the point where the central hydrogen mass fraction has fallen to 0.5, 0.1 and 10$^{-5}$ is shown for each star except the $\chi_0 = 10^{-7}$ case.}
          \label{fig:mesa}
       \end{figure*}

In fig. (\ref{fig:mesa}) we show the Hertsprung-Russell (HR) diagram for stars of one solar mass with initial metallicity $Z=0.02$ (solar metallicity) evolving from the ZAMS to the tip of the red giant branch for $\chi_0=5\times10^{-6},10^{-6},10^{-7}$ and the standard, unmodified case. It is clear from the tracks that stars that are less screened are indeed hotter and more luminous. We can also see that for $\chi_0=10^{-7}$ the tracks are identical along the main sequence but separate in the red-giant phase corresponding to the Newtonian potential dropping due to the increased radius and the star becoming less screened. It appears that the red giant tracks all converge to a similar track. This is due to the fact that the Newtonian potential of these stars is so shallow such that at the mantle they are unscreened to a very high degree, even at $\chi_0=10^{-7}$. In fact, if one examines the tracks in detail then small differences can be discerned.  Note that while the HR tracks for stars of similar masses may possess similar shapes, this does not mean that the other stellar properties are simply a translation in luminosity axis at the same age.

To show this explicitly, we show the star's age and radius at three identical points along the stars main sequence. Notice that it is clear that unscreened stars do indeed have shorter lives than their screened counterparts. It also shows that the radii at the same evolutionary stage tend to be smaller as well. Physically, the extra pressure needed to support the star in modified gravity is produced by increased densities and temperatures over the standard case, demanding a more compact star. We have not shown the values for $\chi_0 = 10^{-7}$ since these stars are almost entirely screened on the main sequence and hence have nearly identical properties to the unmodified stars. Nor have we shown any information about the red giant phase stars. This is because the red giant phase is far shorter than the main sequence and so comparing quantities between stars that are screened to different extents is misleading. Despite the assumptions, it is clear that the luminosity increase described above is still present when more realistic models are used and the missing physics is accounted for. 

\section{Galactic Probes of Modified Gravity with Screening Mechanisms}\label{sec:galaxies}

In this subsection we consider a fully unscreened galaxy composed of main sequence stars and calculate the luminosity compared with an identical unscreened galaxy.\newline

We have seen how the Newtonian potential determines the extent to which an object is screened. Ultimately, we would like to calculate the contribution from the stars to the galactic luminosity in the presence of modified gravity in unscreened galaxies. Treating an individual galaxy as a sphere of mass $M_X$ and radius $R_X$, the characteristic potential is
\begin{equation}\label{eq:sphere_pot}|\Phi_N|\sim \frac{GM_X}{R_X}.\end{equation}Now, dwarf galaxies typically have rotational velocities of the order 50-100 Kms$^{-1}$ and hence, using the virial theorem we have $|GM_X/R_X|\sim\mathrm{10}^{-8}$-10$^{-7}$. Dwarf galaxies are therefore an excellent probe of modified gravity theories with $\chi_0\ge$10$^{-8}$, at least one order of magnitude lower than current experimental bounds. Whereas these dwarf galaxies will not be self-screened, it is entirely possible that dwarf galaxies in clusters may be screened by the Newtonian potential of their neighbours. This will form the basis of any observational method of searching for these theories using stellar effects. 

The number of stars born with mass $M$ within a galaxy is given by the initial mass function (IMF) $\Phi(M)=\dd N/\dd M$, i.e. the number of stars $dN$ within mass range $dM$. In many stellar populations of, where we can resolve individual stars, this relation is empirically found to be roughly universal \footnote{In principle, since the physics of gravitational collapse is expected to be encoded in the IMF, modified gravity may change its functional form. However, we do not attempt to investigate this here since the IMF is empirical and there is no known successful derivation either analytically or numerically of this function from known principles.}. For simplicity, in this work we use the Salpeter IMF \cite{Salpeter:1955it}, $\Phi(M)\propto M^{-2.35}$ with $0.08\le M\le 100 M$; leaving aside questions of whether this function is valid at very high and very low masses. We can then estimate the luminosity increase for an unscreened dwarf galaxy by integrating luminosity (\ref{eq:modlum}) using the fitting functions (\ref{eq:fits}) over the IMF. 

Before doing so however, we must account for the stars that have gone off the main sequence as the IMF only gives the number of stars born. We do this by making use of eqn. (\ref{eq:ms_life}). If the age of the galaxy is $\tau_{\rm age}$ then we assume that all stars with $\tau_{\rm MS}>\tau_{\rm age}$ contribute in their entirety to the luminosity whereas stars whose main sequence lifetimes are less than the age of the galaxy contribute a fraction $\tau_{\rm MS}/\tau_{\rm age}$ of their luminosity. We note that stars that have gone off the main sequence still have a luminosity enhancement in the red giant phase (and beyond), however we do not account for their contribution here. This effect is accounted for in our analysis by including a factor $f_0(M;\tau_{\rm age})$ where
\be
f_0(M;\tau_{\rm age}) = \left\{
  \begin{array}{l l}
   1 & \quad \tau_{\rm MS}>\tau_{\rm age}\\
   \frac{\tau_{\rm MS}}{\tau_{\rm age}} & \quad \tau_{\rm MS}<\tau_{\rm age}\\
  \end{array} \right.
\ee
so that the galactic luminosity is \begin{align}\label{eq:galum}
L_{\mathrm{gal}}(\tau_{\mathrm{age}},&\chi_0) = \\ &\int_{0.08{M_{\odot}}}^{100{M_{\odot}}} dM f_0(M,\tau_{\rm {age}})L(M;\chi_0)\frac{\mathrm{d}N}{\mathrm{d}M}.\end{align}
We can then immediately see that stars whose main sequence lifetime are shorter than the age of the galaxy do not contribute to the luminosity enhancement since the factor of $L(M)^{-1}$ in the main sequence lifetime exactly cancels the factor in the integral. Normalising the integrals so that the total luminous mass of screened and unscreened galaxies are identical (this is required in order to account for the fact that more stars have gone off the main sequence in the unscreened case), we have performed the integral (\ref{eq:galum}) for dwarf galaxies of mass $M_{\rm gal}=$10$^{10}M_\odot$ and age $\tau_{\rm age}=$13 Gyr for $\chi_0$ in the range 10$^{-6}$-10$^{-4}$, a range which, as explained above, is not yet be ruled out observationally. Our results are summarised in table (\ref{tab:luminc}).
\begin{table}[hb]
\centering
\begin{tabular}{|c|c|}
\hline
$\chi_0$ & Luminosity Enhancement \\
\hline
$1$x10$^{-4}$ & 42\% \\
\hline
$1$x10$^{-5}$ & 42\% \\
\hline
$5$x10$^{-6}$ & 29\% \\
\hline
$1$x10$^{-6}$ & 3\% \\
\hline
\end{tabular}
\caption{The luminosity enhancement in unscreened relative to screened dwarf galaxies as a function of $\chi_0$. All values were computed using $\alpha_0 = 1/3$.}
\label{tab:luminc}
\end{table}
Table \ref{tab:luminc} reveals that the enhancement is significant for $\chi_0\gsim$10$^{-6}$, which, as argued above, we would expect. The saturation around $\chi_0\sim$10$^{-5}$ is due to the effect of the decreased main sequence lifetime. If $\chi_0$ is around this value then, as seen in fig. (\ref{fig:lumplot}), the low mass stars are all completely unscreened and so increasing its value further cannot make them any more luminous. Hence, the luminosity enhancement saturates around this value. As noted earlier, red giant stars are still unscreened if $\chi_0\lsim$10$^{-6}$ and these stars tend to be far more luminous than those on the main sequence, see fig. (\ref{fig:mesa}) and so even at these low values of $\chi_0$ it is possible that there will be a significant effect, for example, galaxies that have partially unscreened post main sequence stars and screened main sequence stars will produce different spectra from those where all stars are partially screened. This may then allow observational bounds to be brought down by another order of magnitude or even further.

It is clear that this analysis is only a first approximation. We have not included the contribution from stars in any post main sequence evolutionary phase, nor have we accounted for any galactic effects such as outflows, dust obscuration and supernova feedback.  In addition to this, the Eddington standard morel of main sequence stars do not account for important effects such as convection and metallicities.

Despite these shortcomings, we argue that dwarf galaxies located in intergalactic voids, which are not subject to the gravitational potential of nearby galaxies, should be unscreened whereas those in clusters are not and should therefore exhibit all of the effects described above. Of course one cannot hope to observe two identical galaxies, one screened, the other not, and simply measure the difference in their luminosity. We would, however, expect there to be systematic differences between galaxies located in clusters and those in voids, of which the extent of screening can be estimated, which can be used as a test of these modified theories. By looking for these systematic differences independent constraints on $\chi_0$ in the range 10$^{-4}$-10$^{-6}$ (or possibly lower) can be found. As mentioned above, we also expect the unscreened galaxies to be more metal enriched, and so we expect there to be differences in other galactic properties such as the spectra, colour and magnitude. All of these possibilities are investigated in a future work \cite{future_work} using a complete numerical implementation of both stellar and galactic physics.

\section{conclusion}\label{sec:conclusion}

Scalar-tensor theories of modified gravity with screening mechanisms are an appealing method of accounting for dark energy and the currently unexplained acceleration of the universe whilst still satisfying solar system tests of GR. These theories may be described by two model independent parameters, the enhancement factor of the additional force $\alpha_0$ and $\chi_0$, which parametrises how effectively objects can self-screen. In general, if the Newtonian potential is less than $\chi_0$ then it will be at least partially unscreened. $\alpha_0$ is of $\mathcal{O}(1)$ in any physically interesting theory and most experimental bounds on the theory come from looking at the values of $\chi_0$ that can be tolerated by experimental searches for fifth forces. 

If one demands that the Galaxy is self-screening then one must take a value $\sim 10^{-6}$ but it is possible that our Galaxy is screened by the Local Group and in this case the bound is relaxed to $10^{-4}$. Dwarf galaxies have Newtonian potentials in the range $10^{-7}-10^{-8}$ so that they are unscreened in all models satisfying these bounds; they are hence excellent probes of modified gravity theories. These rough estimates are based on the notion that main sequence stars are sufficiently dense to be self-screened. In the case where the galaxy is screened by the local group the bound of $10^{-4}$ does not come directly from this requirement but from cluster abundances and so tests using astrophysical effects such as those presented in this work could improve this bound.

In this work, we point out that main sequence stars are actually not fully self-screened; sufficiently low mass stars (and stars in the Giant branches) can and do feel the effect of modified gravity, provided that their environment is not screened. Hence, most stars in unscreened galaxies evolved very differently from those in screened galaxies. This means that we can use entire galaxies as probes of modified gravity. This is a particularly crucial point -- galaxies are many orders of magnitude brighter than stars and hence probe a much larger domain of space than stars, opening up the window for testing modified gravity to  much greater length scales and allowing us to access available large data sets on galaxy spectra.

Using a simple analytic model, we have calculated the effect of modified gravity with screening mechanisms on the structure of main sequence stars. There is a significant enhancement in the luminosity in models with $\chi_0\gsim 5\times10^{-6}$. In addition to this, the star's lifetime can be severely reduced, with unscreened stars going off the main sequence more than three times as quickly than an identical screened star. An unscreened dwarf galaxy should then be far more luminous than one which is identical but screened. While this is an idealised situation, there should be systematic offsets between the properties of unscreened dwarf galaxies in voids and those in clusters who are screened by the potential of other galaxies in close proximity. It may then be possible to use this offset as the basis for an observational test of these theories. 

We have estimated this enhancement by summing the luminosity enhancement of the constituent stars in the galaxy, appropriately weighted by the IMF and accounting for stars that have gone off the main sequence since the galaxy was formed. In the most extreme cases, the total enhancement can be up to 40\%. We have not investigated the other effects of the reduced stellar lifetime. Stars that go off the main sequence faster inevitably die faster and so one would expect unscreened dwarf galaxies to have hosted many more generations of stars than identical screened ones of the same age. We therefore expect systematic differences between the spectra and colour \footnote{One could say that chameleons change the colour of galaxies.} in addition to the effects described here. These effects cannot be calculated in our analytic model and we investigate them computationally in a followup work. 

In our analytic treatment, we have made several simplifying assumptions. Firstly, our stellar model does not include the effects of metallicity or nuclear fusion. The stellar luminosity is degenerate with the metallicity and so a full analysis should account for stars of varying metallicity within the galaxy. The lack of nuclear burning prevents us from calculating the dynamical evolution of the model and we therefore do not include the effects of different reaction networks. Since the central temperature rises in order to combat the increased gravitational force, these networks may become significant in stars where they are usually negligible and this has a non-linear effect on the stellar lifetime that we have not accounted for here. Other stellar processes such as convection may be important, however, they are too complicated to include in any analytic model. 

Secondly, our galactic model has not included non-stellar effects such as outflows and dust obscuration, which are likely to be important in determining the luminosity. Nor have we included the effect of post-main sequence stars, which are often more luminous than those on the MS. Since we expect these to be less screened than main sequence stars the effect of neglecting these is, in fact, to underestimate the galactic luminosity. In a follow up work, we shall numerically implement the screening mechanism into computational models that fully account for the effects we have neglected here, some of which we have already presented.\newline

To check that our assumptions do reproduce the gross effects of modified gravity, we implemented the screening mechanism in a full numerical stellar evolution code MESA. The results validated our assumptions and demonstrated the key features of modified gravity in stellar evolution : higher luminosity, shorter lifetimes, smaller radii, and higher effective temperatures. Using these numerical results of stellar evolution, we are constructing realistic galaxies and their spectra with the goal of using them as precision probes of modified gravity.

\begin{acknowledgments} First and foremost, we are incredibly grateful to Bill Paxton for tolerating and answering our many questions about MESA. We thank K. Koyama, A. Pipino, J. Chen, S. Hansen, D. Rudd, J. Zwart and especially L. Hui and P. Chang. ACD, EAL, JAS and DJS acknowledge the STFC, and EAL also acknowledges a NASA Astrophysics grant (09-ATP09-0049).\end{acknowledgments}

\bibliography{ref}

\begin{thebibliography}{25}
\expandafter\ifx\csname natexlab\endcsname\relax\def\natexlab#1{#1}\fi
\expandafter\ifx\csname bibnamefont\endcsname\relax
  \def\bibnamefont#1{#1}\fi
\expandafter\ifx\csname bibfnamefont\endcsname\relax
  \def\bibfnamefont#1{#1}\fi
\expandafter\ifx\csname citenamefont\endcsname\relax
  \def\citenamefont#1{#1}\fi
\expandafter\ifx\csname url\endcsname\relax
  \def\url#1{\texttt{#1}}\fi
\expandafter\ifx\csname urlprefix\endcsname\relax\def\urlprefix{URL }\fi
\providecommand{\bibinfo}[2]{#2}
\providecommand{\eprint}[2][]{\url{#2}}

\bibitem[{\citenamefont{Copeland et~al.}(2006)\citenamefont{Copeland, Sami, and
  Tsujikawa}}]{Copeland:2006wr}
\bibinfo{author}{\bibfnamefont{E.~J.} \bibnamefont{Copeland}},
  \bibinfo{author}{\bibfnamefont{M.}~\bibnamefont{Sami}}, \bibnamefont{and}
  \bibinfo{author}{\bibfnamefont{S.}~\bibnamefont{Tsujikawa}},
  \bibinfo{journal}{Int.J.Mod.Phys.} \textbf{\bibinfo{volume}{D15}},
  \bibinfo{pages}{1753} (\bibinfo{year}{2006}), \eprint{hep-th/0603057}.

\bibitem[{\citenamefont{Weinberg}(1965)}]{Weinberg:1965rz}
\bibinfo{author}{\bibfnamefont{S.}~\bibnamefont{Weinberg}},
  \bibinfo{journal}{Phys.Rev.} \textbf{\bibinfo{volume}{138}},
  \bibinfo{pages}{B988} (\bibinfo{year}{1965}).

\bibitem[{\citenamefont{Will}(2004)}]{Will:2004nx}
\bibinfo{author}{\bibfnamefont{C.}~\bibnamefont{Will}},
  \bibinfo{journal}{Pramana} \textbf{\bibinfo{volume}{63}},
  \bibinfo{pages}{731} (\bibinfo{year}{2004}).

\bibitem[{\citenamefont{Khoury and
  Weltman}(2004{\natexlab{a}})}]{Khoury:2003aq}
\bibinfo{author}{\bibfnamefont{J.}~\bibnamefont{Khoury}} \bibnamefont{and}
  \bibinfo{author}{\bibfnamefont{A.}~\bibnamefont{Weltman}},
  \bibinfo{journal}{Phys.Rev.Lett.} \textbf{\bibinfo{volume}{93}},
  \bibinfo{pages}{171104} (\bibinfo{year}{2004}{\natexlab{a}}),
  \eprint{astro-ph/0309300}.

\bibitem[{\citenamefont{Brax et~al.}(2008)\citenamefont{Brax, van~de Bruck,
  Davis, and Shaw}}]{Brax:2008hh}
\bibinfo{author}{\bibfnamefont{P.}~\bibnamefont{Brax}},
  \bibinfo{author}{\bibfnamefont{C.}~\bibnamefont{van~de Bruck}},
  \bibinfo{author}{\bibfnamefont{A.-C.} \bibnamefont{Davis}}, \bibnamefont{and}
  \bibinfo{author}{\bibfnamefont{D.~J.} \bibnamefont{Shaw}},
  \bibinfo{journal}{Phys.Rev.} \textbf{\bibinfo{volume}{D78}},
  \bibinfo{pages}{104021} (\bibinfo{year}{2008}), \eprint{0806.3415}.

\bibitem[{\citenamefont{Hinterbichler et~al.}(2011)\citenamefont{Hinterbichler,
  Khoury, Levy, and Matas}}]{Hinterbichler:2011ca}
\bibinfo{author}{\bibfnamefont{K.}~\bibnamefont{Hinterbichler}},
  \bibinfo{author}{\bibfnamefont{J.}~\bibnamefont{Khoury}},
  \bibinfo{author}{\bibfnamefont{A.}~\bibnamefont{Levy}}, \bibnamefont{and}
  \bibinfo{author}{\bibfnamefont{A.}~\bibnamefont{Matas}}
  (\bibinfo{year}{2011}), \eprint{1107.2112}.

\bibitem[{\citenamefont{Brax et~al.}(2010{\natexlab{a}})\citenamefont{Brax,
  van~de Bruck, Davis, and Shaw}}]{Brax:2010gi}
\bibinfo{author}{\bibfnamefont{P.}~\bibnamefont{Brax}},
  \bibinfo{author}{\bibfnamefont{C.}~\bibnamefont{van~de Bruck}},
  \bibinfo{author}{\bibfnamefont{A.-C.} \bibnamefont{Davis}}, \bibnamefont{and}
  \bibinfo{author}{\bibfnamefont{D.}~\bibnamefont{Shaw}},
  \bibinfo{journal}{Phys. Rev.} \textbf{\bibinfo{volume}{D82}},
  \bibinfo{pages}{063519} (\bibinfo{year}{2010}{\natexlab{a}}),
  \eprint{1005.3735}.

\bibitem[{\citenamefont{Adams}(2008)}]{Adams:2008ad}
\bibinfo{author}{\bibfnamefont{F.~C.} \bibnamefont{Adams}},
  \bibinfo{journal}{JCAP} \textbf{\bibinfo{volume}{0808}}, \bibinfo{pages}{010}
  (\bibinfo{year}{2008}), \eprint{0807.3697}.

\bibitem[{\citenamefont{Hui et~al.}(2009)\citenamefont{Hui, Nicolis, and
  Stubbs}}]{Hui:2009kc}
\bibinfo{author}{\bibfnamefont{L.}~\bibnamefont{Hui}},
  \bibinfo{author}{\bibfnamefont{A.}~\bibnamefont{Nicolis}}, \bibnamefont{and}
  \bibinfo{author}{\bibfnamefont{C.}~\bibnamefont{Stubbs}},
  \bibinfo{journal}{Phys.Rev.} \textbf{\bibinfo{volume}{D80}},
  \bibinfo{pages}{104002} (\bibinfo{year}{2009}), \eprint{0905.2966}.

\bibitem[{\citenamefont{Chang and Hui}(2010)}]{Chang:2010xh}
\bibinfo{author}{\bibfnamefont{P.}~\bibnamefont{Chang}} \bibnamefont{and}
  \bibinfo{author}{\bibfnamefont{L.}~\bibnamefont{Hui}} (\bibinfo{year}{2010}),
  \eprint{1011.4107}.

\bibitem[{\citenamefont{Davis et~al.}(2011)\citenamefont{Davis, Kotulla, Lim,
  and Sakstein}}]{future_work}
\bibinfo{author}{\bibfnamefont{A.-C.} \bibnamefont{Davis}},
  \bibinfo{author}{\bibfnamefont{R.}~\bibnamefont{Kotulla}},
  \bibinfo{author}{\bibfnamefont{E.~A.} \bibnamefont{Lim}}, \bibnamefont{and}
  \bibinfo{author}{\bibfnamefont{J.}~\bibnamefont{Sakstein}}
  (\bibinfo{year}{2011}).

\bibitem[{\citenamefont{Waterhouse}(2006)}]{Waterhouse:2006wv}
\bibinfo{author}{\bibfnamefont{T.~P.} \bibnamefont{Waterhouse}}
  (\bibinfo{year}{2006}), \eprint{astro-ph/0611816}.

\bibitem[{\citenamefont{Hinterbichler and Khoury}(2010)}]{Hinterbichler:2010es}
\bibinfo{author}{\bibfnamefont{K.}~\bibnamefont{Hinterbichler}}
  \bibnamefont{and} \bibinfo{author}{\bibfnamefont{J.}~\bibnamefont{Khoury}},
  \bibinfo{journal}{Phys. Rev. Lett.} \textbf{\bibinfo{volume}{104}},
  \bibinfo{pages}{231301} (\bibinfo{year}{2010}), \eprint{1001.4525}.

\bibitem[{\citenamefont{Schmidt et~al.}(2009)\citenamefont{Schmidt, Lima,
  Oyaizu, and Hu}}]{Schmidt:2008tn}
\bibinfo{author}{\bibfnamefont{F.}~\bibnamefont{Schmidt}},
  \bibinfo{author}{\bibfnamefont{M.~V.} \bibnamefont{Lima}},
  \bibinfo{author}{\bibfnamefont{H.}~\bibnamefont{Oyaizu}}, \bibnamefont{and}
  \bibinfo{author}{\bibfnamefont{W.}~\bibnamefont{Hu}}, \bibinfo{journal}{Phys.
  Rev.} \textbf{\bibinfo{volume}{D79}}, \bibinfo{pages}{083518}
  (\bibinfo{year}{2009}), \eprint{0812.0545}.

\bibitem[{\citenamefont{Khoury and
  Weltman}(2004{\natexlab{b}})}]{Khoury:2003rn}
\bibinfo{author}{\bibfnamefont{J.}~\bibnamefont{Khoury}} \bibnamefont{and}
  \bibinfo{author}{\bibfnamefont{A.}~\bibnamefont{Weltman}},
  \bibinfo{journal}{Phys. Rev.} \textbf{\bibinfo{volume}{D69}},
  \bibinfo{pages}{044026} (\bibinfo{year}{2004}{\natexlab{b}}),
  \eprint{astro-ph/0309411}.

\bibitem[{\citenamefont{Paxton et~al.}(2011)}]{Paxton:2010ji}
\bibinfo{author}{\bibfnamefont{B.}~\bibnamefont{Paxton}} \bibnamefont{et~al.},
  \bibinfo{journal}{Astrophys. J. Suppl.} \textbf{\bibinfo{volume}{192}},
  \bibinfo{pages}{3} (\bibinfo{year}{2011}), \eprint{1009.1622}.

\bibitem[{\citenamefont{Salpeter}(1955)}]{Salpeter:1955it}
\bibinfo{author}{\bibfnamefont{E.~E.} \bibnamefont{Salpeter}},
  \bibinfo{journal}{Astrophys. J.} \textbf{\bibinfo{volume}{121}},
  \bibinfo{pages}{161} (\bibinfo{year}{1955}).

\bibitem[{\citenamefont{Brax et~al.}(2011)\citenamefont{Brax, Burrage, Davis,
  Seery, and Weltman}}]{Brax:2010uq}
\bibinfo{author}{\bibfnamefont{P.}~\bibnamefont{Brax}},
  \bibinfo{author}{\bibfnamefont{C.}~\bibnamefont{Burrage}},
  \bibinfo{author}{\bibfnamefont{A.-C.} \bibnamefont{Davis}},
  \bibinfo{author}{\bibfnamefont{D.}~\bibnamefont{Seery}}, \bibnamefont{and}
  \bibinfo{author}{\bibfnamefont{A.}~\bibnamefont{Weltman}},
  \bibinfo{journal}{Phys. Lett.} \textbf{\bibinfo{volume}{B699}},
  \bibinfo{pages}{5} (\bibinfo{year}{2011}), \eprint{1010.4536}.

\bibitem[{\citenamefont{Mota and Shaw}(2006)}]{Mota:2006ed}
\bibinfo{author}{\bibfnamefont{D.~F.} \bibnamefont{Mota}} \bibnamefont{and}
  \bibinfo{author}{\bibfnamefont{D.~J.} \bibnamefont{Shaw}},
  \bibinfo{journal}{Phys. Rev. Lett.} \textbf{\bibinfo{volume}{97}},
  \bibinfo{pages}{151102} (\bibinfo{year}{2006}), \eprint{hep-ph/0606204}.

\bibitem[{\citenamefont{Mota and Shaw}(2007)}]{Mota:2006fz}
\bibinfo{author}{\bibfnamefont{D.~F.} \bibnamefont{Mota}} \bibnamefont{and}
  \bibinfo{author}{\bibfnamefont{D.~J.} \bibnamefont{Shaw}},
  \bibinfo{journal}{Phys. Rev.} \textbf{\bibinfo{volume}{D75}},
  \bibinfo{pages}{063501} (\bibinfo{year}{2007}), \eprint{hep-ph/0608078}.

\bibitem[{\citenamefont{Brax et~al.}(2010{\natexlab{b}})\citenamefont{Brax,
  van~de Bruck, Mota, Nunes, and Winther}}]{Brax:2010kv}
\bibinfo{author}{\bibfnamefont{P.}~\bibnamefont{Brax}},
  \bibinfo{author}{\bibfnamefont{C.}~\bibnamefont{van~de Bruck}},
  \bibinfo{author}{\bibfnamefont{D.~F.} \bibnamefont{Mota}},
  \bibinfo{author}{\bibfnamefont{N.~J.} \bibnamefont{Nunes}}, \bibnamefont{and}
  \bibinfo{author}{\bibfnamefont{H.~A.} \bibnamefont{Winther}},
  \bibinfo{journal}{Phys. Rev.} \textbf{\bibinfo{volume}{D82}},
  \bibinfo{pages}{083503} (\bibinfo{year}{2010}{\natexlab{b}}),
  \eprint{1006.2796}.

\bibitem[{\citenamefont{Gannouji et~al.}(2010)}]{Gannouji:2010fc}
\bibinfo{author}{\bibfnamefont{R.}~\bibnamefont{Gannouji}}
  \bibnamefont{et~al.}, \bibinfo{journal}{Phys. Rev.}
  \textbf{\bibinfo{volume}{D82}}, \bibinfo{pages}{124006}
  (\bibinfo{year}{2010}), \eprint{1010.3769}.

\bibitem[{\citenamefont{Mota and Winther}(2010)}]{Mota:2010uy}
\bibinfo{author}{\bibfnamefont{D.~F.} \bibnamefont{Mota}} \bibnamefont{and}
  \bibinfo{author}{\bibfnamefont{H.~A.} \bibnamefont{Winther}}
  (\bibinfo{year}{2010}), \eprint{1010.5650}.

\bibitem[{\citenamefont{Damour and Polyakov}(1994)}]{Damour:1994zq}
\bibinfo{author}{\bibfnamefont{T.}~\bibnamefont{Damour}} \bibnamefont{and}
  \bibinfo{author}{\bibfnamefont{A.~M.} \bibnamefont{Polyakov}},
  \bibinfo{journal}{Nucl. Phys.} \textbf{\bibinfo{volume}{B423}},
  \bibinfo{pages}{532} (\bibinfo{year}{1994}), \eprint{hep-th/9401069}.

\bibitem[{\citenamefont{Gasperini et~al.}(2002)\citenamefont{Gasperini, Piazza,
  and Veneziano}}]{Gasperini:2001pc}
\bibinfo{author}{\bibfnamefont{M.}~\bibnamefont{Gasperini}},
  \bibinfo{author}{\bibfnamefont{F.}~\bibnamefont{Piazza}}, \bibnamefont{and}
  \bibinfo{author}{\bibfnamefont{G.}~\bibnamefont{Veneziano}},
  \bibinfo{journal}{Phys.Rev.} \textbf{\bibinfo{volume}{D65}},
  \bibinfo{pages}{023508} (\bibinfo{year}{2002}), \eprint{gr-qc/0108016}.

\end{thebibliography}

\appendix

\section{Specific Screening Mechanisms}

\subsection{Chameleons}

\begin{figure}
\label{fig:min_cham}
\subfigure[Small {$\rho$}]{\label{fig:small_rho}\includegraphics[width=0.5\textwidth]{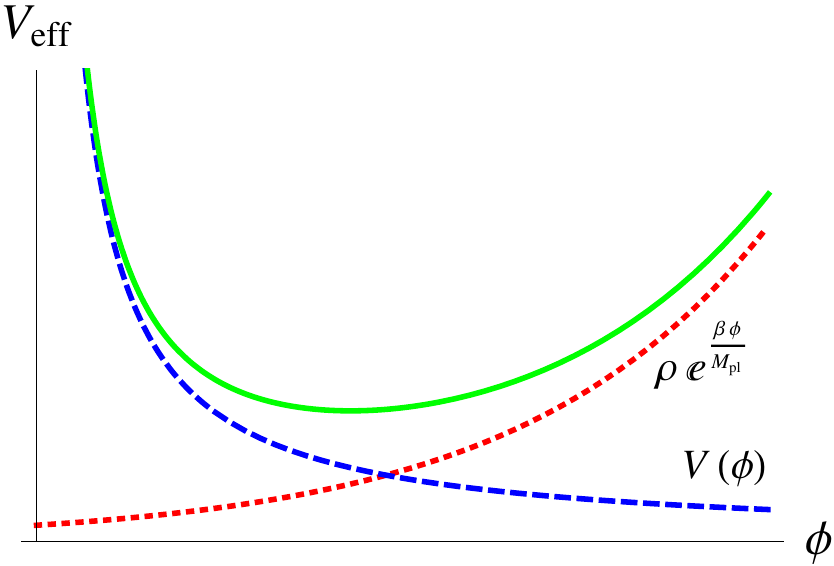}}
\subfigure[Large {$\rho$}]{\label{fig:large_rho}\includegraphics[width=0.5\textwidth]{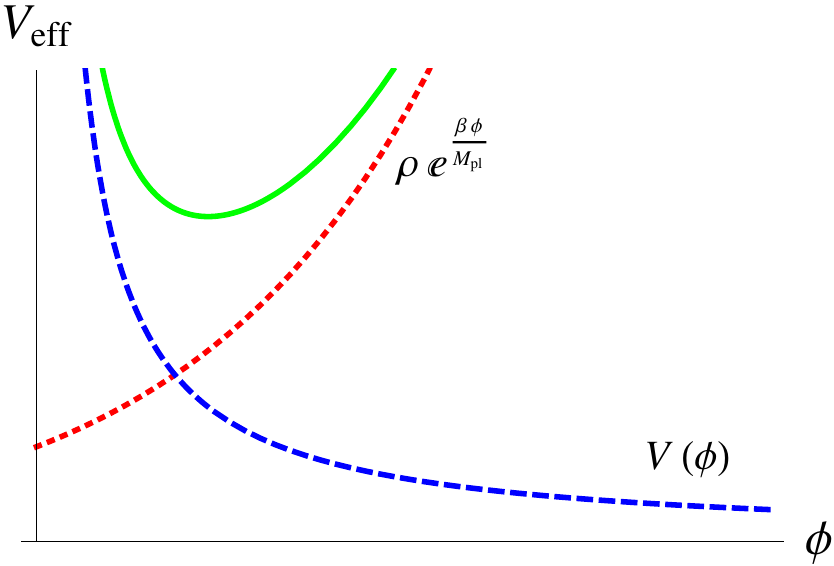}}
\caption{The chameleon effective potential (solid line) for small and large local densities. The dashed lines show the contribution from the potential and the dotted lines show the contribution from the coupling.}
\end{figure}

First proposed by Khoury and Weltman in 2003 \cite{Khoury:2003rn}, chameleon fields typically have a run-away potential such that $V\rightarrow 0$ as $\phi\rightarrow\infty$ (one standard paradigm being $V(\phi)=M^{4+n}/\phi^n$) and a monotonically increasing coupling function of the form
\begin{equation}\label{eq:cham_coupling}A(\phi)=e^{\beta(\phi)/\mpl}.\end{equation}
$\beta$ is usually assumed to be of $\mathcal{O}\!\left(\mpl\right)$, corresponding to a gravitational strength force however $\beta\gg 1$ is not necessarily ruled out by current experimental bounds \cite{Mota:2006ed,Mota:2006fz}. In the standard case $\beta(\phi) = \beta\phi$ where $\beta$ is a constant, however more complicated couplings have been considered \cite{Brax:2010kv,Gannouji:2010fc,Mota:2010uy}. \newline With this choice of potential and coupling function, the field value that minimises the effective potential shifts to smaller field values in high density regions as shown in fig. (\ref{fig:min_cham}). The chameleon screening mechanism relies on the effective mass being much larger in high density environments compared with the cosmological value and hence the fifth force being incredibly short ranged. This can be seen to be the case in fig. (\ref{fig:min_cham}). It is this ``blending in with its environment'' that gives rise to the particles name. In practice, this screening requires the chameleon to sit at the minimum of its effective potential throughout the majority of the interior of the over density and varying only in a thin shell near the surface. This is hence known as \textit{the thin shell effect}.

\subsection{Symmetrons}

Symmetron fields \cite{Hinterbichler:2010es} rely on local symmetry restoration as a method of screening and avoiding fifth force constraints. They are characterised by a $\mathbb{Z}_2$ invariant, symmetry breaking potential \begin{equation}\label{eq:symm_pot}V(\phi)=-\frac{1}{2}\mu^2\phi^2+\frac{1}{4}\lambda\phi^4+\mathcal{O}(\phi^6)\,,\end{equation}
with $\mu>0$ and a coupling (also $\mathbb{Z}_2$ invariant)
\begin{equation}\label{eq:symm_coup}A(\phi)=1+\frac{1}{2M^2}\phi^2+\mathcal{O}(\phi^4)\,,\end{equation}
which are both invariant under the symmetry transformation $\phi\rightarrow -\phi$. We note that the potential and coupling need only take this form in the vicinity of the minimum, and may have a more complicated general form at field values far away from this region. The potential eqn. (\ref{eq:symm_pot}) should hence be viewed as an effective field theory valid at low energies. The requirement that the symmetron force be comparable with that due to gravity in vacuum tunes the parameter $\lambda\sim$10$^{-96}$ and so the effective field theory is valid for all regimes of interest. The effective potential is then
\begin{equation}\label{eq:symm_eff_pot}V_{\mathrm{eff}}(\phi)=\frac{1}{2}\left(\frac{\rho}{M^2}-\mu^2\right)+\frac{1}{4}\lambda\phi^4\end{equation}
and thus the symmetry is restored in sufficiently dense regions such that $\rho>\mu^2 M^2$. In high density regions of space-time the field's vacuum expectation value (VEV), $\phi_{\vev}=0$ whilst in low density regions (i.e. cosmologically), we have
\begin{equation}\label{eq:symm_vev}\phi_{\vev} \approx\pm\frac{\mu}{\sqrt{\lambda}}\,.\end{equation}
The coupling function \ref{eq:symm_coup} then gives the leading order coupling of small perturbations around this VEV to matter as
\begin{equation}\label{eq:symm_matter_coupling}\beta(\phi_{\vev}) = \frac{\mpl\phi_{\vev}}{M^2}.\end{equation}
Evidently, the coupling is proportional the local VEV and is hence zero in dense enough regions such that the symmetry is restored but is strong when the density is small enough that the symmetry is broken. The symmetron parameters are chosen such that the symmetry is broken in our recent history in order to enter the epoch of dark energy domination at a similar time and avoid the coincidence problem. This requires fine-tuning of the mass scales $M$ and $\mu$ such that $H_0^2\!\mpl^2\sim\mu^2M^2$. In addition to this we require $M<10^{-4}\mpl$ in order to satisfy local gravity constraints. Unlike the chameleon and the environmentally dependent dilaton (described below), the symmetron cannot account for the vacuum energy density today. The vacuum energy at the symmetry broken minimum is of order $H_0^2M^2\ll H_0^2\mpl^2$ and so it is necessary to add a constant vacuum energy to the potential in order to reproduce the correct cosmic history. 

\subsection{The Environmentally Dependent Dilaton}

The environmentally dependent dilaton\cite{Brax:2010gi}, is an environmentally dependent generalisation of the Darmour-Polyakov mechanism\cite{Damour:1994zq} that arises from string theory in the strong coupling limit. The starting point is the four dimensional low energy effective action in the string frame
\begin{align}\begin{split}\label{eq:string_frame} S&=\int\dd^4x\sqrt{-\tilde{g}}\left[A(\phi)^-2\frac{\mpl^2\tilde{R}}{2}+\frac{Z(\phi)}{2l_s^2}\tilde{\nabla}_\mu\phi\tilde{\nabla}^\mu\phi-\tilde{V}(\phi)\right]\\ &+S_m[\psi_i,\tilde{g}_{\mu\nu},g_i(\phi)],\end{split}\end{align}
where $\phi$ is the dilaton, $l_s$ is the string length scale ($l_s\sim M_s^{-1}$ with $M_s$ the string scale), and we have allowed the coupling constants to contain some dilaton dependence not induced by the coupling function $A$. It has been argued \cite{Gasperini:2001pc} that the functions appearing in \ref{eq:string_frame} take the following form in the strong coupling limit:
\begin{align}\tilde{V}(\phi)&\sim V_0e^{-\phi}+\mathcal{O}(e^{-2\phi})\\
Z(\phi)&\sim -\frac{2c_1^2}{\lambda^2}+b_Ze^{-\phi}+\mathcal{O}(e^{-2\phi}),\\
\frac{1}{g_i^2}&\sim \frac{1}{\bar{g}_i^{2}}+b_ie^{-\phi}+\mathcal{O}(e^{-2\phi}),
\end{align}with $c_1 = M_s/\mpl\gg1$ and $\lambda \sim\mathcal{O}(1)$-$\mathcal{O}(c_1)$.
Applying this and making the conformal transformation $\tilde{g}_{\mu\nu}=A^2(\phi)g_{\mu\nu}$ we obtain the Einstein frame action
\begin{align}\begin{split}\label{eq:einstein_frame}S&=\int\dd^4x\sqrt{-g}\left[\frac{\mpl^2R}{2}-\mpl^2k^2(\phi)\nabla_\mu\phi\nabla^\mu\phi-V(\phi)\right]\\&+S_m[\Psi_i,g_{\mu\nu},g_i(\phi)],\end{split}\end{align}
where \be\label{eq:k} k^2(\phi)\approx\frac{1}{\lambda^2}+3\beta^2(\phi);\quad \beta(\phi)=\frac{\dd\ln A}{\dd \phi}\ee
and $V(\phi)=A^4\tilde{V}(\phi)$. Note that the field is dimensionless in this model, forcing us to change the definition of $\beta$ relative to the standard case. The non-canonical kinetic term modifies the equations of motion, which (if the contribution from the coupling constants is small) is
\begin{align}\begin{split}\label{eq:eom}\Box\varphi&=\frac{1}{2\mpl^2K(\phi)}\left[-V(\phi)+\beta(\phi)(A(\phi)\rho+V(\phi)\right]\\ &=\frac{1}{2\mpl^2}\frac{\dd V_{\rm eff}}{\dd \varphi}\end{split}\end{align} which then defines the effective potential \be \label{eq:veff}V_{\rm eff}(\varphi)=V_0A^4e^{-\phi}+\rho A(\phi).\ee 
We assume that the coupling function has a minimum at some $\phi=\phi_0$ so that near $\phi_0$ we have
\be\label{eq:dil_coup_min} A(\phi)\approx 1+ \frac{A_2}{2}(\phi-\phi_0)^2\ee with $A_2\gg1$ but $A_2(\phi-\phi_0)^2/2\ll1$. These conditions are required in order to satisfy experimental constraints \cite{Brax:2010gi}. Minimising the effective potential and setting $A(\phi)\approx1$, we have \be \beta(\phi_{\rm min}) = A_2(\phi_{\rm min}-\phi_0)=\frac{V_0e^{-\phi_{\rm min}}}{\rho+4V_0e^{-\phi_{\rm min}}}\ee and thus in high density environments, the coupling function tends to zero, whilst in low density environments $\beta\rightarrow1/4$. Next, we must worry about the strength of the fifth force, which, using the non-canonical expression for the force enhancement is \be \alpha(\phi)=\frac{\lambda^2\beta^2(\phi)}{3\lambda^2\beta^2(\phi)+1},\ee where we have used eqn. (\ref{eq:k}). In this case, the theory is screened in high density environments, $\alpha(\phi_{\rm min})\rightarrow 0$ whilst in the opposite limit $\alpha(\phi)\rightarrow 1/3$ for $\rho\rightarrow 0 $ where minimising the effective potential is equivalent to minimising the potential alone. 

\end{document}